# Communication complexity of approximate Nash equilibria

Yakov Babichenko[*]     Aviad Rubinstein[†]

September 13, 2016

**Abstract**

For a constant $\epsilon$, we prove a $\mathsf{poly}(N)$ lower bound on the (randomized) communication complexity of $\epsilon$-Nash equilibrium in two-player $N \times N$ games. For $n$-player binary-action games we prove an $\exp(n)$ lower bound for the (randomized) communication complexity of $(\epsilon, \epsilon)$-weak approximate Nash equilibrium, which is a profile of mixed actions such that at least $(1 - \epsilon)$-fraction of the players are $\epsilon$-best replying.

## 1   Introduction

Complexity of equilibria has been studied in several complexity models. In particular, computational complexity, query complexity, and communication complexity. Due to recent developments in the field, the computational complexity and the query complexity of approximate Nash equilibria are quite well understood, even for constant approximation value:

- For constant $\epsilon$, there exists a quasi-polynomial algorithm for $\epsilon$-Nash equilibrium in two-player $N \times N$ games [LMM03]. Under the "Exponential Time Hypothesis for PPAD", no better algorithm exists [Rub16]. For better approximation value of $\epsilon = 1/N$ the problem becomes PPAD-complete [DGP09, CDT09].

  For constant $\epsilon$, it is PPAD-hard to compute an $\epsilon$-Nash equilibrium in succinctly representable $n$-player games [Rub14], even for graphical polymatrix binary-action games [Rub15].

- The query complexity of $\epsilon$-Nash equilibrium in two-player $N \times N$ games is $\mathsf{poly}(N)$ [FGGS13]. The query complexity of $\epsilon$-Nash equilibrium in $n$-player binary-action games is $\exp(n)$ [HN13, Bab14, CCT15, Rub16].

The main motivation for this line of research is the insight about the relevance of Nash equilibrium as a predictive solution concept: if specialized algorithms cannot compute an (approximate) equilibrium, it is unreasonable to expect selfish agents to "naturally" converge to one. In the famous words of Kamal Jain, "If your laptop can't find it, then neither can the market". (See also discussions in [DGP09, Nis09b, HM10].) Although extremely useful, lower bounds on computational and query complexity suffer from obvious caveats. Computational complexity lower bounds inherently rely on complexity assumptions (such as $\mathsf{NP} \neq \mathsf{P}$ or PPAD ≠ P); even though these assumptions are widely accepted by computer scientists, they make these theorems less accessible to game theorists and economists. Query complexity lower bounds hold only against a fairly restricted model of accessing data, where the algorithm must pay for querying the utility at each strategy profile; what if, for example, we instead give the algorithm access to a best-response oracle? In particular, no lower bounds were known for convergence to approximate Nash equilibrium via randomized *uncoupled dynamics* (see Subsection 1.1). It is thus of great interest to prove unconditional lower bounds on the much more general model of *communication complexity*, where each player has unrestricted access to information about her own utility.

---

[*]Technion. Email:yakovbab@tx.technion.ac.il. This research was supported by ISF grant 2021296.

[†]UC Berkeley. Email: aviad@eecs.berkeley.edu. Most of this work was done while the author was an intern at Microsoft Research, Hertzeliyah. This research was also partly supported by Microsoft Research PhD Fellowship, NSF grant CCF1408635, and Templeton Foundation grant 3966.

While computational and query complexity of approximate Nash equilibrium are quite well understood, for communication complexity, only results on *pure* Nash equilibrium or *exact* Nash equilibria were known: The communication complexity of pure Nash equilibrium in two-player $N \times N$ game is $\mathsf{poly}(N)$ [CS04], and in $n$-player games it is $\exp(n)$ [HM10]. The communication complexity of exact Nash equilibrium in $n$-player games is also $\exp(n)$ [HM10][1]. No communication complexity lower bounds were known for approximate Nash equilibria. In fact, even for an approximate equilibrium for approximation of value $\epsilon = 1/\mathsf{poly}(N)$, no bounds were known, see [Nis09a]. In this paper we prove the hardness of *approximate* Nash equilibria in the randomized[2] communication complexity model.

**Theorem** (Main Theorem, informal)**.** *There exists a constant $\epsilon > 0$, such that:*

**2-player** *$\epsilon$-Nash equilibrium in two-player $N \times N$ games requires $\mathsf{poly}(N)$ communication.*

**n-player** *$\epsilon$-Nash equilibrium in n-player binary-action games require $2^{\Omega(n)}$ communication. In fact, we prove the exponential lower bound even for a weaker notion of $(\epsilon, \epsilon)$-weak approximate Nash equilibrium, where it is allowed that $\epsilon$-fraction of the players will play an arbitrary action (not necessarily an $\epsilon$-best-reply).*

## 1.1 Uncoupled dynamics

An underling assumption of the Nash equilibrium solution is that players predict correctly the (mixed) action of their opponents (or alternatively predict correctly their expected payoff at each action). One justification for this problematic assumption, which appears in the seminal work of John Nash [Nas51], is that *in some scenarios* players may *learn* the behaviour of their opponents in cases where the game is played repeatedly. This idea led to an extensive study of learning dynamics and their convergence to Nash equilibrium, see e.g. [You04, HMC13, KL93]. One natural, and general, class of adaptive dynamics is that of *uncoupled dynamics* [HMC03, HMC06] where it is assumed that players do not know the utilities of their opponents (but observe their past behaviour). The question on the *existence* of uncoupled dynamics that lead to Nash equilibrium is quite well understood [FY06, HMC06, GL07, Bab12]. Several uncoupled dynamics that converge to approximate Nash equilibrium (or pure Nash equilibrium [You09]) are known. All these dynamics are based on an exhaustive search principle, where at the moment a player realizes she is acting sub-optimally she updates her action to a random one (rather than to an optimal one or a better one). One criticism of these dynamics is that convergence to equilibrium may take

---

[1] The lower bound of [HM10] is exponential in the number of players, but only polynomial in the size of the description of the equilibrium; see [Nis09a].

[2] In an earlier version of this paper we proved hardness for *deterministic* communication. We were able to extend it to randomized communication thanks to a helpful suggestion by Mika Goos [Göö16].

an unreasonably long time in large games where the exhaustive search is done over a large space. This led to the study of the *rate of convergence* of uncoupled dynamics. As pointed out by [CS04] for every solution concept (in particular equilibria solutions), the (randomized) communication complexity of a solution is identical (up to a logarithmic factor) to the rate of convergence by any (randomized) uncoupled dynamics to the solution. This observation initiated the communication complexity study in games. As was mentioned above, the communication complexity, and thus also the rate of convergence of uncoupled dynamics, was known only for exact or pure Nash equilibrium. The question on the rate of convergence of uncoupled dynamics to approximate Nash equilibrium was an open question. Given the fact that all known positive results introduce dynamics that converge to *approximate* Nash equilibrium, this question is central. Our Main Theorem resolves this open question, yielding the following negative result:

**Corollary 1.1** (Uncoupled Dynamics). *There exists a constant $\epsilon > 0$ such that any uncoupled dynamics requires:*

2-**player** *at least* $\mathsf{poly}(N)$ *rounds to converge to an $\epsilon$-Nash equilibrium in two-player $N \times N$ games.*

$n$-**player** *at least* $2^{\Omega(n)}$ *rounds to converge to an $\epsilon$-Nash equilibrium (or even $(\epsilon, \epsilon)$-weak approximate Nash equilibrium) in $n$-player binary-action games.*

## 1.2 Techniques

Proving communication complexity lower bounds for Nash equilibrium is notoriously challenging for two reasons. The first reason, as is common in hardness of Nash equilibrium in other models, is *totality*: there always exists at least one (exact) equilibrium, and the proof of existence induces a non-trivial (yet inefficient) algorithm for finding it. In order to construct hard instances we must carefully hide the equilibrium (we can't just remove it), and make sure that the above algorithm is indeed inefficient for our instances.

Another reason for the communication complexity of approximate equilibrium being an open question for a long time is the fact that there exist efficient *non-deterministic* communication protocols ($\mathsf{polylog}(N)$ for two-player, $\mathsf{poly}(n)$ for $n$-player): verification of equilibrium (exact or approximate) requires only constant communication, and small-representation approximate equilibria always exist (e.g. by [LMM03]). Therefore, the communication complexity lower bounds for approximate equilibria, as we prove in the present paper, show an exponential gap between the non-deterministic and randomized (or even deterministic) communication complexity of a total problem. Such exponential gaps are rare in communication complexity (for exceptions see e.g. [RW90, KRW95, RM97])[3].

In this work, we overcome both obstacles by combining techniques from hardness of Nash equilibrium in other models [HPV89, Shm12, Bab14, Rub16] together with the simulation theorem of [GLM+15]. We note that even given all those techniques, several challenges must be overcome, as is evident by [RW16].

The main steps in our proofs are as follows. First, we prove a *query complexity* hardness result for the problem of finding the end of a line in a particular constant-degree graph. Then we use a *simulation theorem* of [GLM+15] to "lift" this query complexity hardness result to communication complexity hardness. We use a construction of [HPV89, Rub16] to embed this line as a continuous Lipschitz function $f : [0,1]^n \to [0,1]^n$. Finally, we build on ideas from [Shm12, Bab14] to construct a two-player (respectively $n$-player) "imitation game" that simulates both the short communication protocol for the computation of $f$, as well as a fixed-point verification. In particular, every (approximate) Nash equilibrium of the game corresponds to an approximate fixed-point of $f$, which in turn corresponds to an end of a line. Proof overview appears in Section 2.1. The formal proofs appear in Section 4.

---

[3]It is interesting to remark that our result is arguably the first example of a *natural* problem which exhibits such a gap: To the best of our knowledge, approximate Nash equilibrium is the first problem that is not defined in order to exhibit a gap, but rather happens to have one.

## 1.3 Additional related literature

For two-player $N \times N$ games and $\epsilon \approx 0.382$, [CDF+15] show that $\mathsf{polylog}(N)$ communication is sufficient for computing an $\epsilon$-approximate Nash equilibrium (improving over a protocol for $\epsilon \approx 0.438$ due to [GP14]).

For the related notion of correlated equilibrium, in $n$-player games with a constant number of actions, it is known that even exact correlated equilibrium can be computed using only $\mathsf{poly}(n)$-communication, see [HM10, PR08, JLB15]. Interestingly, for exact correlated equilibria, there is an exponential gap between the above communication protocol and the *query complexity* lower bound of [HN13, BB15]. Further discussion on correlated equilibria appears in Section 5.

For the related problem of finding a fixed point, [RW16] study the communication complexity of approximate fixed point of the decomposition. Namely, Alice holds a Lipschitz function $f : A \to B$ Bob holds a Lipschitz function $g : B \to A$, where $A$ and $B$ are compact convex sets, and their goal is to compute a fixed point of the decomposition $g \circ f$. [RW16] prove that the following version of this problem is communicationally hard: find an approximate fixed point of $g \circ f$ on a grid of $A$, when it is promised that such an approximate fixed point *on the grid* exists (the problem is not total).

As discussed earlier, the main motivation for studying the (communication) complexity of Nash equilibrium is understanding its relevance as a predictive solution concept. This is a good place to mention a recent work of Roughgarden [Rou14], which highlights another important motivation for studying the complexity of Nash equilibrium: understanding the *quality* of equilibria. The *Price of Anarchy (POA)* of a game is the ratio between the *social welfare* (sum of players' utilities) in an optimum strategy profile, and the social welfare in the worst Nash equilibrium of that game. Roughgarden [Rou14] provides the following widely applicable recipe for lower bounds on PoA: if a Nash equilibrium can be found efficiently (in particular, via the non-deterministic protocol due to [LMM03]), but approximating the optimal social welfare requires a higher communication complexity (even for non-deterministic protocols, e.g. by reduction from set disjointness), then clearly not all Nash equilibria yield high social welfare.

## 2 Results and proof overview

For two-player games the communication complexity of $\epsilon$-Nash equilibrium is defined to be the problem of finding an $\epsilon$-Nash equilibrium, when Alice holds the utility function of player 1, and Bob holds the utility function of player 2. We prove the following result:

**Theorem 2.1.** *There exists a constant $\epsilon > 0$ such that the randomized communication complexity ($\mathsf{BPP}^{\mathsf{cc}}$) of $\epsilon$-Nash equilibrium in two-player $N \times N$ games is at least $N^\epsilon$.*

For $n$-player games, we consider a two-party communication problem where the set of players $[n]$ is divided into two disjoint subsets $[n] = n^A \uplus n^B$. Alice holds the utilities of the players in $n^A$, and Bob holds the utilities of the players in $n^B$. In particular, this communication problem is easier than the $n$-parties communication problem where each player holds his own utility function. Our negative result holds for the notion of *weak* approximate Nash equilibrium [BPR16], which in particular implies the same negative result for the standard notion of approximate Nash equilibrium (see also Definition 3.3).

**Theorem 2.2.** *There exists a constant $\epsilon > 0$ such that the randomized communication complexity (BPP$^{cc}$) of $(\epsilon, \epsilon)$-weak approximate Nash equilibrium in $n$-player binary-action games is at least $2^{\epsilon n}$.*

The formal proofs appear in Section 4. Below we present the main ideas of the proof.

## 2.1 Overview of the proofs

As mentioned in the Introduction, the proof consists of four main steps. Below we present the ideas of each step.

**Query Complexity of End-of-any-Line**

Our proof starts with the following *query complexity* hardness result (Lemma 4.4): There exists a constant degree graph $G = (V, E)$ with $2^{\Theta(n)}$ vertices, such that finding the end of a line in $G$ requires $2^{\Omega(n)}$ queries. In fact, we prove the hardness result for directed graph $G$ where each vertex has outgoing and incoming degree 2. Therefore, the successor and predecessor of each vertex are binary variables. In particular, for each $v \in V$, the information about its role in the line can be represented using only three bits, which we denote $I(v) \triangleq (T(v), P(v), S(v)) \in \{0, 1\}^3$:

(a) Whether the line goes trough $v$, which is denoted by $T(v)$,

(b) Who is the successor of $v$ (if $v$ in on the line), which is denoted by $S(v)$,

(c) Who is the predecessor of $v$ (if $v$ in on the line), which is denoted by $P(v)$.

**Lemma** (QUERY-EOAL; informal). *Finding an end of any line requires $2^{\Omega(n)}$ queries to $I$.*

Proving hardness for deterministic (or even bounded-error randomized) query complexity is quite easy (see Lemma 4.2). This can be "lifted" to a lower bound on deterministic communication complexity via the simulation theorem of [RM97, GPW15]. In order to eventually obtain hardness for *randomized* communication, we prove the aforementioned query complexity lower bound in the WAPP model [BGM06, GLM$^+$15] (see Section 3.2 for definitions). One challenge is that in the WAPP model, if $I$ describes only one line, finding its endpoint is actually easy. To this end, we consider a collection of (one or two) lines. The proof follows along the lines of [GJPW15, Göö16].

**From Query complexity to Communication Complexity**

We use a *simulation theorem* to "lift" our query complexity lower bound to communication complexity. Obtaining a simulation theorem for the standard randomized model of communication (BPP) is an important open problem (e.g. [GJPW15]). We follow the approach of [GJPW15] (and a wise suggestion from Mika Goos [Göö16]), and circumvent the lack of simulation theorem for BPP by applying the WAPP simulation theorem of [GLM+15] to the WAPP-query complexity hardness we obtained earlier.

The simulated communicationally hard problem has the following form. For each $v \in V$, Alice holds a triplet of vectors $\alpha_{T,v}, \alpha_{S,v}, \alpha_{P,v} \in \{0,1\}^M$ where $M = 2^{O(n)}$, and Bob holds a reasonably small input which is just a triplet of indexes $\beta_{T,v}, \beta_{S,v}, \beta_{P,v} \in [M]$. $T(v)$ is given by the decomposition $T(v) = \alpha_{T,v}(\beta_{T,v})$ (similarly for the successor and predecessor). The simulation theorem of [GLM+15] now implies:

**Corollary** (CC(SIM-EOAL); informal)**.** *Finding an end of any line requires $2^{\Omega(n)}$ bits of communication.*

### Embedding as a continuous function

Our next step is to reduce the problem of finding an end of a line to that of finding a Brouwer fixed point. Here, we use a recent construction of a hard function by [Rub16], which improved over the classic construction of Hirsch et al [HPV89].

We embed the points of the discrete graph $G$ in the continuous space $[-1, 2]^{\Theta(n)}$. Specifically, we embed each vertex $v$ of $G$ into a point $\mathbf{x}_v$ in $[-1, 2]^{\Theta(n)}$ and each edge $(v, w)$ in $G$ into a (continuous) path in $[-1, 2]^{\Theta(n)}$ that connects the corresponding points $\mathbf{x}_v$ and $\mathbf{x}_w$. Roughly speaking, [Rub16] shows that we can map a collection of lines in $G$ into a Lipschitz function $f : [-1, 2]^{\Theta(n)} \to [-1, 2]^{\Theta(n)}$ such that:

1. The computation of $f$ can be done using *local information about $I$*. Namely, for points that are close to $\mathbf{x}_v$ it is sufficient to know $I(v)$ to compute $f$. For points that are close to a path that corresponds to the edge $(v, w)$ but far from the points $\mathbf{x}_v, \mathbf{x}_w$ it is sufficient to know whether $(v, w)$ is an edge in the line (in particular, knowing either $I(u)$ or $I(v)$ suffices). For points that are far from all paths $(v, w)$, $f$ does not depend on $I$ at all (thus can be computed without any communication).

2. Any (normalized) $\|\cdot\|_2$-approximate fixed point of $f$ can be mapped (efficiently) back to an end of some line in $I$.

Property 1 induces the following efficient communication protocol for computing $f(\mathbf{x})$: Bob finds $v$ such that $\mathbf{x}$ is close to $\mathbf{x}_v$, and sends $\beta_{T,v}, \beta_{S,v}, \beta_{T,v}$; Alice replies with $I(v) = \left(\alpha_{T,v}(\beta_{T,v}), \alpha_{T,v}(\beta_{T,v}), \alpha_{T,v}(\beta_{T,v})\right)$, and they each use $I(v)$ to locally compute $f(\mathbf{x})$. (Similarly, if $\mathbf{x}$ is close to the path corresponding to edge $(v, w)$, they use a similar protocol to compute $I(v)$ and $I(w)$.)

By Property 2, we have:

**Corollary** (CC(BROUWER); informal)**.** *Finding a (normalized) $\|\cdot\|_2$-approximate fixed point of $f$ requires $2^{\Omega(n)}$ bits of communication.*

### Two-player game

Naively thinking, we would like to design a game where Alice chooses a point $\mathbf{x} \in [-1, 2]^{\Theta(n)}$ (on the $\varepsilon$-grid) and Bob chooses a point $\mathbf{y} \in [-1, 2]^{\Theta(n)}$ (on the $\varepsilon$-grid). Alice's utility will be given by $-\|\mathbf{x} - \mathbf{y}\|_2^2$, and Bob's utility will be given by[4] $-\|\mathbf{y} - f(\mathbf{x})\|_2^2$. Then, by applying similar arguments to those in [Shm12, Bab14, Rub16] we can deduce that every approximate Nash equilibrium corresponds to an approximate fixed point, and thus also to an end of a line.

---
[4]Note that here it is crucial that we use the normalized $\|\cdot\|_2$ to obtain payoffs bounded in $[-9, 0]$; using the non-normalized $\|\cdot\|_2$ we get payoffs in $[-\sqrt{n}, 0]$.

However, the above idea is obviously incorrect because Bob's utility depends on $f$, whereas in the communication problem his utility should depend on the $\beta$s only. Our key idea is to use the fact that $f$ can be computed locally to design a somewhat similar game where similar phenomena to those in the "naive" game will occur in approximate equilibria.

Bob doesn't know $f$, but to compute $f(\mathbf{x})$ he should only know the local information about the vertex (or vertices) that correspond to $\mathbf{x}$. We incentivize Alice and Bob to combine their private information about the corresponding vertex (or vertices) by the following utilities structure.

- Alice's first component of utility is given by $-\|\mathbf{x}-\mathbf{y}\|_2^2$. As in the "naive" game, in any approximate Nash equilibrium Alice will play points in the $\epsilon$-cube of the $\epsilon$-grid that contains $\mathsf{E}[\mathbf{y}]$ with probability close to one.

- Bob tries to guess the vertex $v$ (or the vertices $v, w$) that correspond to the point $\mathbf{x}$. Since $\mathbf{x}$ (almost always) belongs to the same $\epsilon$-cube, in any (approximate) Nash equilibrium, his guess should be correct (with high probability). In addition, Bob should announce the $\beta$ indexes $\beta_T, \beta_S$ and $\beta_P$ of $v$ (of $v$ and $w$). Again, we incentivize him to do so by defining that he should "guess" also these $\beta$ indexes, and in an (approximate) equilibrium his guess should be correct (w.h.p).

- We want Alice to announce $I(v)$ (similarly for $w$ in case of two vertices). Thus, we ask her to guess the decomposition $\alpha_{v^B}(\beta^B)$ where $v^B$ and $\beta^B$ are the announced $v$ and $\beta$ by Bob. In (approximate) equilibrium, since Bob announces the correct $v$ and $\beta$ (w.h.p), this incentivizes her to announce the correct $I(v)$ (w.h.p).

- Now Bob uses the local information of $I(v)$ (and $I(w)$) to compute $f$. Namely, his last utility component is defined by $-\left\|\mathbf{y} - f_{I^A(v), I^A(w)}(\mathbf{x})\right\|_2^2$ where $f_{I^A(v), I^A(w)}$ is Bob's "estimation" of $f$ under the relevant local information announced by Alice. In (approximate) equilibrium Alice announces the correct local information (w.h.p), thus Bob computes $f$ correctly (w.h.p).

Summarizing, the (approximate) equilibrium analysis of the presented game is similar to the analysis of the naive game, because in (approximate) equilibrium $f$ is computed correctly (w.h.p). But unlike the naive game, here Alice's utility depends only on the $\alpha$s and Bob's utility only on the $\beta$s.

### $n$-player game: $\epsilon$-WSNE

The $n$-player game reduction is based on the same ideas as the two-player reduction. For clarity, we present first the idea of a reduction that proves the following weaker result:

There exists a constant $\epsilon > 0$ such that the communication complexity of $\epsilon$-well supported approximate Nash equilibrium in $n$-player games with constant number of actions for each player is at least $2^{cn}$ for some constant $c$.

After that, we explain how we can strengthen this result in two aspects: first to improve the constant-number-of-action to binary-action, second to improve the $\epsilon$-well supported Nash equilibrium to $(\epsilon, \epsilon)$-weak approximate equilibrium.

The idea is to replace a single player- Alice- who chooses $\mathbf{x}$ in the $\epsilon$-grid of $[-1,2]^{\Theta(n)}$ by a population of $\Theta(n)$ players $\{p_{x_i}\}_{i \in \Theta(n)}$; each player $p_{x_i}$ in the population is responsible for the $i$th coordinate of $\mathbf{x}$. The payoff of player $p_{x_i}$ is given by $-|x_i - y_i|^2$. This incentivizes player $p_{x_i}$ to play either a single, or two adjacent actions, in the $\epsilon$-grid of the segment $[-1,2]$ (in every WSNE). By looking at the action profile of all $p_{x_i}$ players we get the same phenomenon as in the two-player case: every point $x$ in the support of Alice's players belongs to the same $\epsilon$-cube of the $\epsilon$-grid.

Now, we replace the guess of $v \in \{0,1\}^{\Theta(n)}$, that is done by Bob, by population of size $\Theta(n)$ where again each player is responsible to a single coordinate of $v$. Again in a WSNE all players will guess correctly.

Similarly for the guess of $\beta$: we think of $\beta \in [M]^3$ as an element of $\{0,1\}^{3 \log M}$ and we construct a population of $3 \log M$ players, each controls a single bit.

Similarly for Alice's guesses of $I^A(v)$ and $I^A(v)$: we construct 6 players, each chooses a bit.

Finally, we again replace the choice of $\mathbf{y} \in [-1,2]^{\Theta(n)}$ by a population of $\Theta(n)$ players $p_{y_i}$. Each is responsible to a single coordinate. The payoff of player $p_{y_i}$ is given by $-|y_i - [f_{I^A(v),I^A(w)}(\mathbf{x})]_i|^2$. The analysis of this game is very similar to the two-player game analysis.

### $n$-player game: $(\epsilon, \epsilon)$-Weak ANE and binary actions

In the above reduction, the $\mathbf{x}$-type (and $\mathbf{y}$-type players) have $3/\epsilon$ actions each. To construct a binary action game we use the technique of [Bab14]. We replace each such player by a population of $3/\epsilon$ players, each is located at a point in the $\epsilon$-grid of the segment $[-1,2]$. Player that is located at $j \in [-1,2]$ (on the $\epsilon$-grid) has to choose between the two points $j$ or $j + \epsilon$. In a WSNE all players are located from the left of $y_i$ will choose $j + \epsilon$, and all players are located from the right of $y_i$ will choose $j$.

More tricky, is to generalize this reduction to weak approximate equilibria. Recall that in weak approximate equilibria, a constant fraction of players may play an arbitrary suboptimal action. Here we take into account both,

1. Players that are not $\epsilon$-best replying, and

2. Players that are $\epsilon$-best replying, but put small positive weight on the inferior action (among the two) and the realization of their mixed action turned out to be the inferior action.

In order to be immune from these, irrational, small constant fraction of players, we use error correcting codes[5]. Let $E_\beta: \{0,1\}^{3 \log M} \to \{0,1\}^{\Theta(3 \log M)}$ be a good binary error correcting code. Instead of having a population of size $3 \log M$ which tries to guess $\beta$, we replace it by a population of size $\Theta(3 \log M)$ where each player tries to guess his bit in the *encoding* of $\beta$. Now even if a small constant fraction of players will act irrationally, the *decoding* of the action profile of the $\beta$-type players will turn out to be $\beta$. We use the same idea for all types of populations ($\mathbf{x}$-type, $\mathbf{y}$-type, $v$-type and $I$-type). This idea completes the reduction for weak approximate equilibria.

---

[5]In fact, we use error correcting codes even earlier, in [Rub16]'s modification construction of hard Brouwer function.

# 3 Preliminaries

**Notation** We use $\mathbf{0}_n$ (respectively $\mathbf{1}_n$) to denote the length-$n$ vectors whose value is 0 (1) in every coordinate. For vectors $\mathbf{x}, \mathbf{y} \in \mathbb{R}^n$, we let $\|\mathbf{x} - \mathbf{y}\|_2 \triangleq \sqrt{\frac{1}{n} \sum_{i \in [n]} (x_i - y_i)^2}$ denote the *normalized* 2-norm. Unless specified otherwise, when we say that $\mathbf{x}$ and $\mathbf{y}$ are $\Delta$-close (or $\Delta$-far), we mean $\Delta$-close in normalized 2-norm.

## 3.1 Different notions of approximate Nash equilibrium

A mixed strategy of player $i$ is a distribution $x_i$ over $i$'s set of actions, $A_i$. We say that a vector of mixed strategies $\mathbf{x} \in \times_j \Delta A_j$ is a *Nash equilibrium* if every strategy $a_i$ in the support of every $x_i$ is a best response to the actions of the mixed strategies of the rest of the players, $x_{-i}$. Formally,

$$\forall a_i \in \mathrm{supp}\,(x_i) \quad \mathsf{E}_{a_{-i} \sim x_{-i}} \left[ u_i(a_i, a_{-i}) \right] = \max_{a' \in A_i} \mathsf{E}_{a_{-i} \sim x_{-i}} \left[ u_i(a', a_{-i}) \right].$$

Equivalently, $\mathbf{x}$ is a Nash equilibrium if each mixed strategy $x_i$ is a best response to $x_{-i}$:

$$\mathsf{E}_{\mathbf{a} \sim \mathbf{x}} \left[ u_i(\mathbf{a}) \right] = \max_{x'_i \in \Delta A_i} \mathsf{E}_{\mathbf{a} \sim (x'_i; x_{-i})} \left[ u_i(\mathbf{a}) \right].$$

Each of those equivalent definitions can be generalized to include approximation in a different way.

**Definition 3.1** ($\epsilon$-Approximate Nash Equilibrium). We say that $\mathbf{x}$ is an $\epsilon$-*Approximate Nash Equilibrium* ($\epsilon$-*ANE*) if each $x_i$ is an $\epsilon$-best response to $x_{-i}$:

$$\mathsf{E}_{\mathbf{a} \sim \mathbf{x}} \left[ u_i(\mathbf{a}) \right] \geq \max_{x'_i \in \Delta A_i} \mathsf{E}_{\mathbf{a} \sim (x'_i; x_{-i})} \left[ u_i(\mathbf{a}) \right] - \epsilon.$$

On the other hand, we generalize the first definition of Nash equilibrium in the following stricter definition:

**Definition 3.2** ($\epsilon$-Well-Supported Nash Equilibrium). $\mathbf{x}$ is a $\epsilon$-*Well-Supported Nash Equilibrium* ($\epsilon$-*WSNE*) if every $a_i$ in the support of $x_i$ is an $\epsilon$-best response to $x_{-i}$:

$$\forall a_i \in \mathrm{supp}\,(x_i) \quad \mathsf{E}_{a_{-i} \sim x_{-i}} \left[ u_i(a_i, a_{-i}) \right] \geq \max_{a' \in A_i} \mathsf{E}_{a_{-i} \sim x_{-i}} \left[ u_i(a', a_{-i}) \right] - \epsilon.$$

**WeakNash**

We can further relax the (already more lenient) notion of $\epsilon$-ANE by requiring that the $\epsilon$-best response condition only hold for most of the players (rather than all of them).

**Definition 3.3** (($\epsilon, \delta$)-WeakNash [BPR16]). We say that $\mathbf{x}$ is an $(\epsilon, \delta)$-*WeakNash* if for a $(1 - \delta)$-fraction of $i$'s, $x_i$ is an $\epsilon$-best mixed response to $x_{-i}$:

$$\Pr_i \left[ \mathsf{E}_{\mathbf{a} \sim \mathbf{x}} \left[ u_i(\mathbf{a}) \right] \geq \max_{x'_i \in \Delta A_i} \mathsf{E}_{\mathbf{a} \sim (x'_i; x_{-i})} \left[ u_i(\mathbf{a}) \right] - \epsilon \right] \geq 1 - \delta.$$

## 3.2 WAPP and Conical Juntas

The class WAPP (Weak Almost-wide PP) was first introduced in [BGM06] in the context of computational complexity. More recently, it has inspired the following models of query and communication complexity:

**Definition 3.4** (WAPP; [GLM+15])**.** The $\mathsf{WAPP}_\epsilon$ {query / communication} complexity of a problem is given by the minimum complexity[6] of an {algorithm / protocol} that, for some parameter $\gamma = \gamma(n)$:

**Completeness** On all 1-instances, output 1 with probability $\in [(1-\epsilon)\gamma, \gamma]$;

**Soundness** On all 0-instances, output 1 with probability $\in [0, \epsilon\gamma]$.

For proving lower bounds against algorithms in the WAPP query model, we will use the following observation due to Goos et al:

**Lemma 3.5** (WAPP decision trees are conical juntas, [GLM+15, Fact 29 of full version])**.** *Without loss of generality, for each outcome $r$ of the randomness, the* WAPP *query algorithm non-adaptively queries a single conjunction $h^r$ and returns the value of this conjunction.*

## 3.3 Simulation Theorems

Let $D : \{0,1\}^N \to \{0,1\}$ be a decision problem. We consider the following query complexity models. Each query is an index $k \in [N]$ and the answer is the $k$-th bit of the input.

- The deterministic query complexity of $D$, denoted by $\mathsf{P}^{\mathrm{dt}}(D)$.

- The $\mathsf{WAPP}_\delta$ query complexity of $D$, denoted by $\mathsf{WAPP}^{\mathrm{dt}}_\delta(D)$.

We also consider the following communication complexity models. Here, for every $k \in [N]$ Alice holds a vector $\alpha_k \in \{0,1\}^M$ and Bob holds an index $\beta_k \in [M]$, for some $M = \mathsf{poly}(N)$. Their goal is to compute $D$ for the input $(\alpha_1(\beta_1), \ldots, \alpha_N(\beta_N))$.

- The two-party deterministic communication complexity of the simulated problem $D$, denoted by $\mathsf{P}^{\mathrm{cc}}(\text{Sim-}D)$.

- The two-party $\mathsf{WAPP}_\epsilon$ communication complexity of the simulated problem $D$, denoted by $\mathsf{WAPP}^{\mathrm{cc}}_\epsilon(\text{Sim-}D)$.

- The (more standard) bounded error two-party probabilistic communication complexity of the simulated problem $D$, denoted by $\mathsf{BPP}^{\mathrm{cc}}(\text{Sim-}D)$.

It is easy to see that for any constant $\delta$ and problem $D$,

$$\mathsf{BPP}^{\mathrm{cc}}(\text{Sim-}D) = \Omega\left(\mathsf{WAPP}^{\mathrm{cc}}_\delta(\text{Sim-}D)\right).$$

To "lift" from query complexity hardness to communication complexity, we use the following *simulation theorem* for WAPP, due to Goos et al [GLM+15].

---

[6]Public-coin communication protocols also incur an additional penalty of $\log(1/\gamma)$.

**Theorem 3.6** (WAPP Simulation Theorem, [GLM+15, Theorem 2][7])**.** *There exists* $M = \mathsf{poly}(N)$ *such that for any constants* $0 < \epsilon < \delta < 1/2$, $\mathsf{WAPP}^{\mathsf{cc}}_\epsilon(\mathrm{SIM}\text{-}D) = \Omega\left(\mathsf{WAPP}^{\mathsf{dt}}_\delta(D)(\log N)\right)$.

For ease of presentation, let us also mention the following simulation theorem for deterministic algorithms/protocols; it is due to Goos et al [GPW15], which in turn is based on the work of Raz and McKenzie [RM97].

**Theorem 3.7** (Deterministic Simulation Theorem, [GPW15])**.** *There exists* $M = \mathsf{poly}(N)$ *such that* $\mathsf{P}^{\mathsf{cc}}(\mathrm{SIM}\text{-}D) = \Theta\left(\mathsf{P}^{\mathsf{dt}}(D)(\log N)\right)$.

## 4 Proofs

In Section 4.1 we prove a deterministic query lower bound for (a special case of) the end-of-any-line problem. In Section 4.2 we prove a query lower bound for the end-of-any-line problem in the WAPP model. In Section 4.3 we show how the lower bounds of Sections 4.1 and 4.2 can be "lifted" to get a hard problem in the deterministic and randomized (respectively) communication complexity models. We note that the communicationally hard problem is the same problem for the deterministic and the randomized models. Obviously, the randomized communication complexity result is stronger. Nevertheless, we present here the proof of both results because the deterministic model is easier to follow.

In Sections 4.4, 4.5, and 4.6 we reduce the communicationally hard end-of-any-line problem to the approximate Nash equilibrium problem. We note that this reduction holds for both the deterministic and randomized communication complexity models.

### 4.1 A deterministic query complexity lower bound

Let $G$ be a *directed* graph with the vertices $V = \{0,1\}^n \times \{0,1\}^n \times [n+1]$. Each vertex $(v_1, v_2, k)$, where $v_1, v_2 \in \{0,1\}^n$ and $k \in [n]$, has two outgoing edges to the vertices $(v_1, v_2^{k+1}(0), k+1)$ and $(v_1, v_2^{k_1}(1), k+1)$, where $v^j(0) = (v_1, \ldots, v_{j-1}, 0, v_{j+1}, \ldots, v_n)$. We call $(v_1, v_2^{k+1}(0), k+1)$ the *0-successor of* $v$, and $(v_1, v_2^{k+1}(1), k+1)$ the *1-successor of* $v$. Each vertex $v = (v_1, v_2, n+1)$ has a single outgoing edge to the vertex $(v_2, v_1, 0)$. Note that the incoming degree of each vertex $v = (v_1, v_2, k) \in V$ is at most two. For $k = 1$ there is a single incoming edge from $(v_2, v_1, n+1)$. For $k > 1$ there are two incoming edges from $(v_1, v_2^k(0), k-1)$ and from $(v_1, v_2^k(1), k-1)$. We call $(v_1, v_2^k(0), k-1)$ the *0-predecessor of* $v$, and $(v_1, v_2^k(1), k-1)$ the *1-predecessor of* $v$.

We define the QUERY END-OF-THE-LINE (QUERY-EOTL) to be the problem of finding the end of a line in $G$ that starts at the point $\mathbf{0}_{2n+1}$. More formally, we represent a line in $G$ by a triple $I(v) \triangleq (T(v), S(v), P(v))$ where $T(v) \in \{0,1\}$ indicates whether the line goes through $v$, $S(v) \in \{0,1\}$ indicates who is the successor of $v$, and $P(v) \in \{0,1\}$ indicates who is the predecessor of $v$ (here we use the fact that each vertex has outgoing and incoming degree of at most two). Throughout the paper, we slightly abuse notation and use $S(v)/P(v)$ to refer both to the bits, and to the corresponding vertices (i.e. the $S(v)/P(v)$-successor/predecessor of $v$). The end of the line is the vertex $v^*$ such that $T(v^*) = 1$ but $T(S(v^*)) = 0$.

**Definition 4.1.** The problem QUERY-EOTL is given by
**Input:** A line $I = (T, S, P)$ over the graph $G$ that starts at the point $\mathbf{0}_{2n+1}$.

---
[7]The statement in [GLM+15] is in fact stronger in the sense that rather than giving Bob a pointer $\beta_i$ to an arbitrary vector $\alpha_i$ held by Alice, it suffices to use a simpler and more compact inner-product gadget. Furthermore, the inner product gadget can be replaced by any two-source extractor.

**Output:** The first bit ($[v^*]_1$) of the end of the line vertex.
**Queries:** Each query is a vertex $v \in V$. The answer is the triplet of bits $I(v) = (T(v), S(v), P(v)) \in \{0,1\}^3$.

**Lemma 4.2** (Deterministic query complexity). $\mathsf{P}^{\mathrm{dt}}(\textsc{Query-EotL}) = \Omega(2^n)$.

*Proof.* We choose a permutation $\pi$ over $\{0,1\}^n \setminus \{\mathbf{0}_n\}$ uniformly at random, and set $\pi(0) \triangleq \mathbf{0}_n$. $\pi$ induces a line of length $\Theta(2^n \cdot n)$ over $G$, starting at $\mathbf{0}_{2n+1}$, ending at $(\pi(2^n-1), \pi(2^n-1), 0)$, and where where two consecutive vertices $v = \pi(i)$ and $w = \pi(i+1)$ are mapped to the following line of $n+1$ edges:

$$(v, v, 0) \to \cdots \to (v, (w_{[1,k]}, v_{[k+1,n]}), k) \to \cdots \to (v, w, n) \to (w, w, 0). \tag{1}$$

Here $(w_{[1,k]}, v_{[k+1,n]})$ denotes the vector with first $k$ coordinates as in $w$ and the last $n-k$ coordinates as in $v$.

The information of a single query of Query-EotL (for the above class of lines) can be extracted from $\pi(i-1), \pi(i)$ and $\pi(i+1)$. Therefore Query-EotL is at least as hard as the problem of finding the first bit of the last element in a random permutation, where each query returns the previous, the current, and the next vertices. Conditioning on the answers to $k$ queries $\pi(q_1-1), \pi(q_1), \pi(q_1+1), \ldots, \pi(q_k-1), \pi(q_k), \pi(q_k+1)$, the last element of the permutation is still uniformly random across all vertices that are not $\pi(q_1), \ldots, \pi(q_k), \pi(q_1-1), \ldots, \pi(q_k-1), \pi(q_1+1), \ldots, \pi(q_k+1)$. This proves that the latter problem requires $\Omega(2^n)$ queries. $\square$

## 4.2 A stronger (WAPP) query complexity lower bound

We consider the same graph $G$ over $\Theta(2^n \cdot n)$ vertices described earlier. But now we construct either one line as before, or two vertex-disjoint lines, one of which again starts at the point $\mathbf{0}_{2n+1}$, and the other at an arbitrary point. The goal is to find an end or starting point of either line, other than the trivial solution $\mathbf{0}_{2n+1}$. More precisely, we will consider instances that satisfy the promise that all non-trivial end and starting points have the same first bit, and the algorithm is expected to return this bit.

We call our new problem the Query End-of-any-Line (Query-EoaL) to be the problem of finding a non-trivial end end of a line. More formally, we represent a collection of (at most two) lines in $G$ by a triple $I(v) \triangleq (T(v), S(v), P(v))$ where $T(v) \in \{0,1\}$ indicates whether a line goes through $v$, $S(v) \in \{0,1\}$ indicates who is the successor of $v$, and $P(v) \in \{0,1\}$ indicates who is the predecessor of $v$ (here we use the fact that each vertex has outgoing and incoming degree of at most two). At the end of each line there is a vertex $v^*$ such that $T(v^*) = 1$ but $T(S(v^*)) = 0$. Similarly, at the beginning of each line, there is a vertex $v^*$ such that $T(v^*) = 1$ but $T(P(v^*)) = 0$ (where for one line $v^* = \mathbf{0}_{2n+1}$).

**Definition 4.3.** The problem Query-EoaL is given by
**Input:** A collection of vertex-disjoint lines $I = (T, S, P)$ over the graph $G$, one of which starts at the point $\mathbf{0}_{2n+1}$.
**Output:** The first bit ($[v^*]_1$) of any end or starting vertex $v^* \neq \mathbf{0}_{2n+1}$. (We will only consider instances where this first bit is the same for all $v^* \neq \mathbf{0}_{2n+1}$.)
**Queries:** Each query is a vertex $v \in V$. The answer is the triplet of bits $I(v) = (T(v), S(v), P(v)) \in \{0,1\}^3$.

(Notice that Query-EoaL is a generalization of Query-EotL.)

**Lemma 4.4** (WAPP query complexity). $\mathsf{WAPP}^{\mathrm{dt}}_{0.1}(\textsc{Query-EoaL}) = \Omega(2^n)$.

*Proof.* Assume by contradiction that there is a $\mathsf{WAPP}_{0.1}$ algorithm with query complexity $o(2^n)$. We assume without loss of generality, that whenever the algorithm queries a vertex of the form $(v, (w_{[1,k]}, v_{[k+1,n]}), k)$ (for some $v, w \in \{0,1\}^n$ and $k \in \{0, \ldots, n\}$), it also tests whether $(v, v, 0)$ is the end of a line by querying $(v, v, 0)$ and $S(v, v, 0)$ (note that this only incurs a constant blowup in the query complexity).

By Lemma 3.5, we can also assume without loss of generality that the algorithm takes the following form: For each random string $r$, it non-adaptively queries $I$ on a subset $U^r \subset V$ of size $|U| = o(2^n)$ and compares it to a guess $I^r(u)$ for each $u \in U$. It returns 1 if and only if $I^r(u) = I(u)$ for all $u \in U^r$. Let $C(r, I)$ denote the function that takes a random string $r$ and a collection of vertex-disjiont lines $I$ and outputs a bit in $\{0, 1\}$ as above. As we mentioned above, we require that if $(v, (w_{[1,k]}, v_{[k+1,n]}), k) \in U^r$, then so are $(v, v, 0)$ and $S(v, v, 0)$.

We consider inputs drawn from three distributions:

- $X$: Draw a permutation $\pi$ over $\{0, 1\}^n \setminus \{\mathbf{0}_n\}$ uniformly at random conditioning on the first bit of the last vertex satisfying $\left[\pi(2^n - 1)\right]_1 = 1$. Set $\pi(0) \triangleq \{\mathbf{0}_n\}$, and let $I$ be the line induced from $\pi$, starting at $\mathbf{0}_{2n+1}$, ending at $(\pi(2^n - 1), \pi(2^n - 1), 0)$, and defined as in (1).

- $Y$: Draw $\pi$ as in $X$. "Cut" $\pi$ into two by picking a random pair[8] $(u, w)$ of consecutive vertices in $\pi$ whose first bits are both 1, and disconnecting them. I.e. we now let $I$ represent two lines: one from $\mathbf{0}_{2n+1}$ to $(u, u, 0)$, and the other from $(w, w, 0)$ to $(\pi(2^n - 1), \pi(2^n - 1), 0)$. (Where each line is again defined according to (1).)

- $Z$: Draw $\pi$ conditioning on the first bit of the last vertex satisfying $\left[\pi(2^n-1)\right]_1 = 0$. Let $I$ be the line induced from $\pi$, starting at $\mathbf{0}_{2n+1}$ and ending at $(\pi(2^n - 1), \pi(2^n - 1), 0)$.

By the requirement from the $\mathsf{WAPP}_{0.1}$ algorithm, we must have that for some $\gamma > 0$,

$$\Pr_r \Pr_{I \sim X}[C(r, I) = 1] \in [0.9\gamma, \gamma] \tag{2}$$

$$\Pr_r \Pr_{I \sim Y}[C(r, I) = 1] \in [0.9\gamma, \gamma] \tag{3}$$

$$\Pr_r \Pr_{I \sim Z}[C(r, I) = 1] \in [0, 0.1\gamma]. \tag{4}$$

For each $i = 0, 1, \ldots$, let $A_i$ denote the set of $r$'s such that $I^r$ expects to observe $i$ ends of lines. I.e. for $i$ distinct $v_1, \ldots, v_i \in \{0, 1\}^n$, $T^r(v_j, v_j, 0) = 1$ and $T^r(S^r(v_j, v_j, 0)) = 0$.

First observe that for $i \geq 2$, random strings in $A_i$ do not contribute at all to (2) because $I$ never agrees with $I^r$. We therefore have:

$$\Pr_r \Pr_{I \sim X}[C(r, I) = 1] = \Pr_r \Pr_{I \sim X}[C(r, I) = 1 \text{ AND } r \in A_0] + \Pr_r \Pr_{I \sim X}[C(r, I) = 1 \text{ AND } r \in A_1]. \tag{5}$$

Now, for $A_0$, after observing $o(2^n)$ queries $(U^r, I^r)$, one has only a negligible advantage at distinguishing between $X$ and $Z$ (see also proof of Lemma 4.4). Therefore,

$$\Pr_r \Pr_{I \sim Z}[C(r, I) = 1 \text{ AND } r \in A_0] \geq (1 - o(1)) \Pr_r \Pr_{I \sim X}[C(r, I) = 1 \text{ AND } r \in A_0]. \tag{6}$$

---
[8]We condition on such a pair being present. With high probability there are $\Omega(2^n)$ such pairs, even after any fixing of $o(2^n)$ positions of the permutation.

Similarly, for $A_1$, $I^r$ has almost *double* the chance of agreeing with $I \sim Y$ than $I \sim X$ since with $o(2^n)$ queries it's impossible to distinguish between the end of the line starting at $\mathbf{0}_{2n+1}$ and the end of the spurious line. ($I^r$ may expect to observe the second starting point, but that also does not contribute to (2).) We therefore have,

$$\Pr_r \Pr_{I \sim Y}[C(r,I) = 1 \text{ AND } r \in A_1] \geq (2 - o(1)) \Pr_r \Pr_{I \sim X}[C(r,I) = 1 \text{ AND } r \in A_1]. \tag{7}$$

Plugging (6) and (7) into (5) we have that,

$$2 \Pr_r \Pr_{I \sim Z}[C(r,I) = 1] + \Pr_r \Pr_{I \sim Y}[C(r,I) = 1] \geq (2 - o(1)) \Pr_r \Pr_{I \sim X}[C(r,I) = 1]$$

$$\geq 1.7\gamma \tag{Ineq. (2)},$$

which is a contradiction to the (weighted) sum of (3) and (4). □

### 4.3 Communicationally hard, discrete end-of-any-line problem

In order to use a simulation theorem (Theorem 3.7 for deterministic and Theorem 3.6 for randomized communication complexity), we define the following simulation variant of QUERY-EOAL:

**Definition 4.5** (SIM-EOAL). We let $N = 2^n \cdot 2^n \cdot (n+1) \cdot 3$.
**Input:** For each $v \in \{0,1\}^n \times \{0,1\}^n \times [n+1]$, Alice receives three vectors $\alpha_v^T, \alpha_v^S, \alpha_v^P \in \{0,1\}^M$, and Bob receives three indices $\beta_v^T, \beta_v^S, \beta_v^P \in [M]$.
   We define

$$T(v) = \alpha_v^T(\beta_v^T), S(v) = \alpha_v^S(\beta_v^S), \text{ and } P(v) = \alpha_v^P(\beta_v^P). \tag{8}$$

We simulate the problem QUERY-EOAL, therefore we restrict attention to inputs such that $(T, S, P)$ that are defined in (8) meet all the requirements of QUERY-EOAL.
   **Output:** The first bit $([v^*]_1)$ of a non-trivial end or start of a line $(v^*, v^*, 0) \neq \mathbf{0}_{2n+1}$.

Applying the deterministic Simulation Theorem (Theorem 3.7) to the query complexity lower bound (Lemma 4.2) gives a lower bound on the deterministic communication complexity of a discrete end of line problem SIM-EOAL.

**Corollary 4.6.** $\mathsf{P}^{\mathsf{cc}}(\text{SIM-EOAL}) = \Omega(2^n)$.

Applying the randomized WAPP Simulation Theorem (Theorem 3.6) to the query complexity lower bound (Lemma 4.4) gives a stronger lower bound on the randomized communication complexity of a discrete end of line problem SIM-EOAL.

**Corollary 4.7.** $\mathsf{BPP}^{\mathsf{cc}}(\text{SIM-EOAL}) = \Omega(\mathsf{WAPP}^{\mathsf{cc}}_{0.01}(\text{SIM-EOAL})) = \Omega(2^n)$.

### 4.4 Embedding a line as a local Lipschitz function

It will be more convenient to define $G$ as a graph over $\{0,1\}^{2n+\log(n+1)}$.
   Let $m = \Theta(2n + \log(n+1)) = \Theta(n)$ and let $E:\{0,1\}^{2n+\log(n+1)} \to \{0,1\}^m$ denote the encoding function of a good binary error correcting code. We embed the discrete graph $G$ into the continuous cube $[-1,2]^{4m}$.

The vertex $v$ is embedded to the point $(E(v), E(v), \mathbf{0}_m, \mathbf{0}_m) \in [-1,2]^{4m}$, which is called *the embedded vertex*.

For two vertices $v, w \in V$ such that $(v, w)$ is an edge in the graph $G$, we define five vertices:

$$\mathbf{x}^1(v, w) \triangleq (E(v), E(v), \mathbf{0}_m, \mathbf{0}_m)$$
$$\mathbf{x}^2(v, w) \triangleq (E(v), E(v), \mathbf{1}_m, \mathbf{0}_m)$$
$$\mathbf{x}^3(v, w) \triangleq (E(v), E(w), \mathbf{1}_m, \mathbf{0}_m)$$
$$\mathbf{x}^4(v, w) \triangleq (E(v), E(w), \mathbf{0}_m, \mathbf{0}_m)$$
$$\mathbf{x}^5(v, w) \triangleq (E(w), E(w), \mathbf{0}_m, \mathbf{0}_m).$$

Note that $\mathbf{x}^1(v, w)$ is the embedded vertex $v$, $\mathbf{x}^5(v, w)$ is the embedded vertex $w$.

The line that connects the points $\mathbf{x}^i(v, w)$ and $\mathbf{x}^{i+1}(v, w)$ is called a *Brouwer line segment*. The union of these four Brouwer line segments is called the *embedded edge* $(v, w)$. It is not hard to check that non-consecutive Brouwer line segments are $\Omega(1)$-far one from the other, and in particular it implies that non-consecutive embedded edges are sufficiently far one from the other.

The following Proposition shows that the end-of-any-line problem can be reduced to the problem of finding an approximate fixed point of a continuous Lipschitz function, when the function is "local" in the following sense: every edge in $G$ is embedded as a path in the continuous hypercube (as described above). For points close to the embedding of an edge, $f$ depends only on the "local behaviour" of the lines $I$ at the endpoints of this edge; for all other points, $f$ is independent of the lines $I$.

**Proposition 4.8** (Essentially [Rub16]). *There exist constants $\delta, h > 0$ such that given a line $I = (T, S, P)$ over $G$ there exists a function $f = f(I) =: [-1,2]^{4m} \to [-1,2]^{4m}$ with the following properties:*

1. $\|f(\mathbf{x}) - \mathbf{x}\|_2 > \delta$ *for every $\mathbf{x}$ that in not $h$-close to the embedded edge of any non-trivial end or start of a line (i.e., the embedding of the edge $(P(v^*), v^*)$ such that $T(v^*) = 1$ but $T(S(v^*)) = 0$; or the edge $(v^*, S(v^*))$ for $v^*$ such that $T(P(v^*)) = 0$, $T(v^*) = 1$, and $v^* \neq \mathbf{0}_{2n+1}$).*

2. $f$ *is $O(1)$-Lipschitz in $\|\cdot\|_2$ norm.*

3. $f$ *is local in the sense that it can be defined as an interpolation between a few (in fact, 64) functions, $\{f_{I_1, I_2} : [-1,2]^{4m} \to [-1,2]^{4m}\}_{I_i \in \{0,1\}^3}$, that do not depend on the lines $I$ and such that:*

    (a) *If the first $m$-tuple of coordinates of $\mathbf{x}$ is $6h$-close to the encoded vertex $E(v)$, but the second $m$-tuple of coordinates of $\mathbf{x}$ is $6h$-far from any encoded vertex $E(w)$ then $f_{I(v), I_2}(\mathbf{x}) = f(\mathbf{x})$ for every $I_2 \in \{0,1\}^3$.*

    (b) *If the second $m$-tuple of coordinates of $\mathbf{x}$ is $6h$-close to the encoded vertex $E(w)$, but the first $m$-tuple of coordinates of $\mathbf{x}$ is $6h$-far from any encoded vertex $E(v)$ then $f_{I_1, I(w)}(\mathbf{x}) = f(\mathbf{x})$ for every $I_1 \in \{0,1\}^3$.*

    (c) *If the first $m$-tuple of coordinates of $\mathbf{x}$ is $6h$-close to the encoded vertex $E(v)$, and the second $m$-tuple of coordinates of $\mathbf{x}$ is $6h$-close to the encoded vertex $E(w)$ then $f_{(I(v), I(w))}(\mathbf{x}) = f(\mathbf{x})$.*

(d) *If none of the above conditions are satisfied, then $f_{I_1,I_2}(\mathbf{x}) = f(\mathbf{x})$ for every $I_1, I_2 \in \{0,1\}^3$.*

A very similar proposition was recently proved in [Rub16]. Property (3) in Proposition 4.8 differs a little from the way "locality" is formalized in [Rub16], but it is an immediate consequence of the construction. For completeness, we present the proof of Proposition 4.8 in Appendix A.

## 4.5 Two-Player game

**Theorem** (Theorem 2.1, restated). *There exists a constant $\epsilon > 0$ such that the communication complexity of $\epsilon$-Nash equilibrium in two-player $N \times N$ games is at least $N^\epsilon$.*

We construct a two-player game between Alice and Bob of size $N_A \times N_B$ for

$$N_A \triangleq (3/\epsilon)^{4m} \cdot 2^3 = 2^{\Theta(n)}$$
$$N_B \triangleq (3/\epsilon)^{4m} \cdot (2^{2n+\log(n+1)})^2 \cdot M^3 = 2^{\Theta(n)}.$$

such that Alice's utility depends on $\{\alpha_v^T, \alpha_v^S, \alpha_v^P\}_v$ only, Bob's utility depends on $\{\beta_v^T, \beta_v^S, \beta_v^P\}_v$ only, and all $\epsilon^4$-approximate Nash equilibria of the game correspond to a $\delta$-fixed point of $f$ from Proposition 4.8. By property 1 in Proposition 4.8, any fixed point of $f$ corresponds to a non-trivial end or start of a line in $I$.

### 4.5.1 The game

In this subsection we construct our reduction from SIM-EOAL to the problem of finding an $\epsilon$-WSNE.

**Strategies**

Recall that $\delta$ is the desired approximation parameter for Brouwer fixed point in the construction of Proposition 4.8. We let $\epsilon$ be a sufficiently small constant; in particular, $\epsilon = O(\delta)$ (this will be important later for Inequality (17)).

Each of Alice's actions corresponds to an ordered tuple $(\mathbf{x}, I_v^A, I_w^A)$, where:

- $\mathbf{x} \in [-1, 2]^{4m}$, where the interval $[-1, 2]$ is discretized into $\{-1, -1 + \epsilon, \ldots, 2 - \epsilon, 2\}$;
- $I_v^A \triangleq (t_v^A, s_v^A, p_v^A) \in \{0,1\}^3$ and $I_w^A \triangleq (t_w^A, s_w^A, p_w^A) \in \{0,1\}^3$.

Each of Bob's actions corresponds to an ordered tuple $(\mathbf{y}, v^B, w^B, \beta_v^B, \beta_w^B)$, where:

- $\mathbf{y} \in [-1, 2]^{4m}$, where the interval $[-1, 2]$ is discretized into $\{-1, -1 + \epsilon, \ldots, 2 - \epsilon, 2\}$;
- $v^B, w^B \in \{0,1\}^{2n+\log(n+1)}$ are vertices in the graph $G$.
- $\beta_v^B = (\beta_v^{B,T}, \beta_v^{B,S}, \beta_v^{B,P}) \in [M]^3$ and $\beta_w^B = (\beta_w^{B,T}, \beta_w^{B,S}, \beta_w^{B,P}) \in [M]^3$ are triples of indexes.

**Utilities**

Alice's and Bob's utilities decompose as

$$U^A \triangleq U^A_{\text{Imitation}} + U^A_{\text{GuessV}} + U^A_{\text{GuessW}}.$$
$$U^B \triangleq U^B_{\text{Brouwer}} + U^B_{\text{GuessV}} + U^B_{\text{GuessW}}.$$

The first component of Alice's utility depends only on the first components of her and Bob's strategies; it is given by:

$$U^A_{\text{Imitation}}(\mathbf{x}; \mathbf{y}) \triangleq -\|\mathbf{x} - \mathbf{y}\|_2^2.$$

Given the first component $\mathbf{x} \in [-1, 2]^{4m}$ of Alice's strategy, we define two decoding functions $D_v, D_w : [-1, 2]^{4m} \to \{0, 1\}^n$ as follows. Let $R_v(\mathbf{x}) \in \{0, 1\}^m$ be the rounding of the first $m$-tuple of coordinates of $\mathbf{x}$ to $\{0, 1\}^m$; let $D_v(\mathbf{x}) = E^{-1}(R_v(\mathbf{x})) \in \{0, 1\}^{2n + \log(n+1)}$, where $E^{-1}$ denote the decoding of the error correcting code from Section 4.4. We define $D_w(\mathbf{x}) \in \{0, 1\}^{2n + \log(n+1)}$ analogously with respect to the second $m$-tuple of coordinates of $\mathbf{x}$. The second components of Bob's utility is now given by $U^B_{\text{GuessV}} = 1$ iff he guesses correctly the vertex $D_v(\mathbf{x})$, and the corresponding $\beta$ operation on this vertex. Namely,

$$U^B_{\text{GuessV}}(v^B, \beta_v^B; \mathbf{x}) = \begin{cases} 1 & \text{if } v^B = D_v(\mathbf{x}) \text{ and } \beta_v^B = (\beta_{D_v(\mathbf{x})}^T, \beta_{D_v(\mathbf{x})}^S, \beta_{D_v(\mathbf{x})}^P) \\ 0 & \text{otherwise.} \end{cases}$$

Similarly we define Bob's third component $U^B_{\text{GuessW}}$ with respect to $D_w(\mathbf{x})$.

Note that Bob knows the indexes $\beta_v^T, \beta_v^S, \beta_v^P$ (for every $v$), thus to achieve $U^B_{\text{Guess}} = 1$ Bob needs to guess correctly only the vertices $D_v(\mathbf{x}), D_w(\mathbf{x})$ and announce the corresponding triplet of $\beta$ indexes.

Going back to Alice, the second component of her utility is given by $U^A_{\text{GuessV}} = 1$ iff she guesses correctly the triplet $I(v^B) = (T(v^B), S(v^B), P(v^B))$ when the calculation of $T, S, P$ is done by the decomposition of $\alpha(\beta^B)$. Namely,

$$U^A_{\text{GuessV}}(I_v^A; v^B, \beta^B) = \begin{cases} 1 & \text{if } I_v^A = (\alpha_{v^B}^T(\beta_v^{B,T}), \alpha_{v^B}^S(\beta_v^{B,S}), \alpha_{v^B}^P(\beta_v^{B,P})) \\ 0 & \text{otherwise.} \end{cases}$$

Similarly we define Alice's third component $U^B_{\text{GuessW}}$.

Finally, the first component of Bob's utility is given by:

$$U^B_{\text{Brouwer}}(\mathbf{y}; \mathbf{x}, e^A) \triangleq -\left\|f_{I_v^A, I_w^A}(\mathbf{x}) - \mathbf{y}\right\|_2^2.$$

where the function $f_{I_1, I_2}$ is defined in Proposition 4.8.

### 4.5.2 Analysis of game

In this subsection, we prove the reduction from Sim-EoAL to finding an $\epsilon^4$-ANE. The proof proceeds via a sequence of lemmas that establish the structure of any $\epsilon^4$-ANE.

**Lemma 4.9.** *In every $\epsilon^4$-ANE $(\mathcal{A}; \mathcal{B})$, it holds that $\|\mathbf{x} - \mathsf{E}_{\mathbf{y} \sim \mathcal{B}}[\mathbf{y}]\|_2^2 = O(\epsilon^2)$ with probability of at least $1 - \epsilon^2$ (where the probability is taken over $\mathcal{A}$).*

*Proof.* We denote $\mathsf{E}_i(\mathcal{B}) = \mathsf{E}_{\mathbf{y} \sim \mathcal{B}}[y_i]$, $\mathsf{E}(\mathcal{B}) = (\mathsf{E}_1(\mathcal{B}), \ldots, \mathsf{E}_n(\mathcal{B}))$ is the vector of expectations, and $\mathrm{Var}(\mathcal{B}) = (\mathrm{Var}_{\mathbf{y} \sim \mathcal{B}}[y_1], \ldots, \mathrm{Var}_{\mathbf{y} \sim \mathcal{B}}[y_n])$ is the vector of variances. For every $\mathbf{x}$ we can rewrite

$$\begin{aligned} U^A_{\text{IMITATION}}(\mathbf{x}, \mathcal{B}) &= -\mathsf{E}_{\mathbf{y} \sim \mathcal{B}} \|\mathbf{x} - \mathbf{y}\|_2^2 \\ &= -\frac{1}{4m} \sum_{i \in [4m]} \mathsf{E}_{\mathbf{y} \sim \mathcal{B}}\left[(x_i - y_i)^2\right] \\ &= -\frac{1}{4m} \sum_{i \in [4m]} \left[(x_i - y_i(\mathcal{B}))^2 + \mathrm{Var}_{\mathbf{y} \sim \mathcal{B}}[y_i]\right] \\ &= -\|\mathbf{x} - \mathsf{E}(\mathcal{B})\|_2^2 - \|\mathrm{Var}(\mathcal{B})\|_2^2. \end{aligned} \qquad (9)$$

Since the variance of the $y_i$'s, as well as $U^A_{\text{GUESSV}}$ and $U^A_{\text{GUESSW}}$, do not depend on $\mathbf{x}$, Alice's best response to $\mathcal{B}$ is

$$\mathbf{x}^* = ([\mathsf{E}_1(\mathcal{B})]_\epsilon, \ldots, [\mathsf{E}_n(\mathcal{B})]_\epsilon)$$

where $[\cdot]_\epsilon$ denotes the rounding to the closest $\epsilon$ integer multiplication. $\mathbf{x}^*$ yields a payoff of at least

$$U^A_{\text{IMITATION}}(\mathbf{x}^*, \mathcal{B}) \geq -\frac{\epsilon^2}{4} - \|\mathrm{Var}(\mathcal{B})\|_2^2.$$

Note that in every $\epsilon^4$-ANE Alice assigns a probability of at most $1 - \epsilon^2$ to actions that are $\epsilon^2$-far from optimal. By Equation (9) this implies that the probability of Alice to choose a vector $\mathbf{x}$ that satisfies $\|\mathbf{x} - \mathsf{E}(\mathcal{B})\|_2^2 \geq \epsilon^2 + \frac{\epsilon^2}{4}$ is at most $\epsilon^2$. □

**Lemma 4.10.** *In every $\epsilon^4$-ANE $(\mathcal{A}; \mathcal{B})$, if the first $m$-tuple of coordinates of $\mathsf{E}_{\mathbf{y} \sim \mathcal{B}}[\mathbf{y}]$ is $6h$-close to the binary encoding $E(v)$ of a vertex $v$, then*

$$v^B = v, \text{ and } \beta_v^B = (\beta_v^T, \beta_v^S, \beta_v^P) \qquad (10)$$

*with probability of at least $1 - O(\epsilon^4)$ (where the probability is taken over $\mathcal{B}$).*

*Proof.* By Lemma 4.9 and the triangle inequality, with probability of at least $1 - \epsilon^2$, the first $m$-tuple of $\mathbf{x}$ is $O(h)$-close to $E(v)$. Rounding to $R_v(\mathbf{x}) \in \{0,1\}^m$ can at most double the distance to $E(v)$ in each coordinate. Therefore, the Hamming distance of $R_v(\mathbf{x})$ and $E(v)$ is $O(h)$. Hence $R_v(\mathbf{x})$ is correctly decoded as $D_v(\mathbf{x}) = v$, with probability of at least $1 - \epsilon^2$.

Since $v^B, \beta_v^B$ do not affect $U^B_{\text{BROUWER}} + U^B_{\text{GUESSW}}$, Bob's utility from guessing $v^B = v$, and $\beta_v^B = (\beta_v^T, \beta_v^S, \beta_v^P)$ is at least $1 - \epsilon^2$. Whereas his utility from guessing any other guess is at most $\epsilon^2$. Therefore, Bob assigns probability at least $1 - \epsilon^4/(1 - 2\epsilon^2)$ to actions that satisfy (10). □

A similar lemma holds for the second $m$-tuple of $\mathbf{x}$ and the vertex $w$:

**Lemma 4.11.** *In every $\epsilon^4$-ANE $(\mathcal{A}; \mathcal{B})$, if the second $m$-tuple of coordinates of $\mathsf{E}_{\mathbf{y} \sim \mathcal{B}}[\mathbf{y}]$ is $6h$-close to the binary encoding $E(W)$ of a vertex $w$, then*

$$w^B = w, \text{ and } \beta_w^B = (\beta_w^T, \beta_w^S, \beta_w^P)$$

*with probability of at least $1 - O(\epsilon^4)$ (where the probability is taken over $\mathcal{B}$).*

Since Alice receives the correct $v^B$ and $\beta^B$, we also have:

**Lemma 4.12.** *In every $\epsilon^4$-ANE $(\mathcal{A}; \mathcal{B})$, if the first m-tuple of coordinates of $\mathsf{E}_{\mathbf{y} \sim \mathcal{B}}[\mathbf{y}]$ is 6h-close to the binary encoding $E(v)$ of a vertex $v$, then*

$$I_v^A = (\alpha_v^T(\beta_v^T), \alpha_v^S(\beta_v^S), \alpha_v^P(\beta_v^P))$$

*with probability $1 - O(\epsilon^4)$ (where the probability is taken over $\mathcal{A}$ and $\mathcal{B}$).*

*Proof.* Follows immediately from Lemma 4.10 and the fact that $I_v^A$ does not affect $U_{\text{IMITATION}}^A + U_{\text{GUESSW}}^A$. □

A similar lemma holds for the second $m$-tuple of $\mathbf{x}$ and the vertex $w$:

**Lemma 4.13.** *In every $\epsilon^4$-ANE $(\mathcal{A}; \mathcal{B})$, if the second m-tuple of coordinates of $\mathsf{E}_{\mathbf{y} \sim \mathcal{B}}[\mathbf{y}]$ is 6h-close to the binary encoding $E(W)$ of a vertex $w$, then*

$$I_w^A = (\alpha_w^T(\beta_w^T), \alpha_w^S(\beta_w^S), \alpha_w^P(\beta_w^P))$$

*with probability $1 - O(\epsilon^4)$ (where the probability is taken over $\mathcal{A}$ and $\mathcal{B}$).*

**Lemma 4.14.** *In every $\epsilon^4$-ANE $(\mathcal{A}; \mathcal{B})$, $f_{I_v^A, I_w^A}(\mathbf{x}) = f(\mathbf{x})$ with probability $1 - O(\epsilon^2)$.*

*Proof.* Follows immediately from Lemmas 4.12 and 4.13 and the "locality" condition in Proposition 4.8. □

The following corollary completes the analysis of the 2-player game.

**Corollary 4.15.** *In every $\epsilon^4$-ANE $(\mathcal{A}; \mathcal{B})$, $\|\mathsf{E}_{\mathbf{x}' \sim \mathcal{A}}[\mathbf{x}'] - f(\mathsf{E}_{\mathbf{x}' \sim \mathcal{A}}[\mathbf{x}'])\|_2 < \delta$.*

*Proof.* We recall that in Lemma 4.9 we have proved that

$$\|\mathbf{x} - \mathsf{E}_{\mathbf{y} \sim \mathcal{B}}[\mathbf{y}]\|_2^2 = O(\epsilon^2) \tag{11}$$

with probability $1 - O(\epsilon^2)$. This also implies that $\mathbf{x}$ is, with high probability, close to its expectation:

$$\begin{aligned}
\|\mathbf{x} - \mathsf{E}_{\mathbf{x}' \sim \mathcal{A}}[\mathbf{x}']\|_2^2 &\leq \left(\|\mathbf{x} - \mathsf{E}_{\mathbf{y} \sim \mathcal{B}}[\mathbf{y}]\|_2 + \|\mathsf{E}_{\mathbf{x}' \sim \mathcal{A}}[\mathbf{x}'] - \mathsf{E}_{\mathbf{y} \sim \mathcal{B}}[\mathbf{y}]\|_2\right)^2 & \text{(Triangle inequaltiy)} \\
&\leq 2\|\mathbf{x} - \mathsf{E}_{\mathbf{y} \sim \mathcal{B}}[\mathbf{y}]\|_2^2 + 2\|\mathsf{E}_{\mathbf{x}' \sim \mathcal{A}}[\mathbf{x}'] - \mathsf{E}_{\mathbf{y} \sim \mathcal{B}}[\mathbf{y}]\|_2^2 & \text{(AM-GM inequaltiy)} \\
&\leq 2\|\mathbf{x} - \mathsf{E}_{\mathbf{y} \sim \mathcal{B}}[\mathbf{y}]\|_2^2 + 2\mathsf{E}_{\mathbf{x}' \sim \mathcal{A}}\left[\|\mathbf{x}' - \mathsf{E}_{\mathbf{y} \sim \mathcal{B}}[\mathbf{y}]\|_2^2\right] & \text{(Convexity of } \|\cdot\|_2^2\text{)} \\
&= O(\epsilon^2), & \text{(Lemma 4.9)} \tag{12}
\end{aligned}$$

with probability $1 - O(\epsilon^2)$.

Using that $f$ is $O(1)$-Lipschitz together with Equation (12), we get that

$$\|f(\mathbf{x}) - f(\mathsf{E}_{\mathbf{x}' \sim \mathcal{A}}[\mathbf{x}'])\|_2^2 = O(\epsilon^2) \tag{13}$$

with probability $1 - O(\epsilon^2)$.

By Lemma 4.14 we know that $f_{I_v^A, I_w^A}(\mathbf{x}) = f(\mathbf{x})$ with probability $1 - O(\epsilon^2)$, which implies that

$$\|\mathsf{E}_{\mathbf{x}' \sim \mathcal{A}}[f_{I_v^A, I_w^A}(\mathbf{x}')] - \mathsf{E}_{\mathbf{x}' \sim \mathcal{A}}[f(\mathbf{x}')]\|_2^2 = O(\epsilon^2). \tag{14}$$

Using similar arguments to those of Lemma 4.9 we can show that

$$\left\|\mathbf{y} - \mathsf{E}_{\mathbf{x}' \sim \mathcal{A}}[f_{I_v^A, I_w^A}(\mathbf{x}')]\right\|_2^2 = O(\epsilon^2) \tag{15}$$

with probability $1 - O(\epsilon^2)$. As in the derivation of Equation (12), this implies:

$$\left\|\mathbf{y} - \mathsf{E}_{\mathbf{y}' \sim \mathcal{B}}[\mathbf{y}']\right\|_2^2 = O(\epsilon^2) \tag{16}$$

with probability $1 - O(\epsilon^2)$.

With probability $1 - O(\epsilon^2)$ Inequalities (12),(11),(16),(15),(14),(13) hold simultaneously. In such a case, by the triangle inequality and by applying the inequalities in the exact above order, we have

$$\left\|\mathsf{E}_{\mathbf{x}' \sim \mathcal{A}}[\mathbf{x}'] - f(\mathsf{E}_{\mathbf{x}' \sim \mathcal{A}}[\mathbf{x}'])\right\|_2^2 = O(\epsilon^2) < \delta^2. \tag{17}$$

□

*Proof of Theorem 2.1.* Any communication protocol that solves the $\epsilon^4$-Nash equilibrium problem in games of size $N \times N$ for $N = 2^{\Theta(n)}$ induces a communication protocol for the problem SIM-EOAL: Alice constructs her utility in the above presented game using her private information of the $\alpha$s, Bob constructs his utility using the $\beta$s. They implement the communication protocol to find an $\epsilon^4$-Nash equilibrium, and then both of them know $\mathsf{E}_{\mathbf{x} \sim \mathcal{A}}[\mathbf{x}]$ which is a $\delta$-approximate fixed point of $f$ (by Corollary 4.15). Using $D_v$ they decode the vertex $v^*$ and they know the first coordinate of $v^*$.

Using Corollary 4.7 we deduce that the communication complexity of $\epsilon^4$-Nash equilibrium in games of size $2^{\Theta(n)} \times 2^{\Theta(n)}$ is at least $2^{\Omega(n)}$. □

### 4.6 n-player game

**Theorem** (Theorem 2.2, restated). *There exists a constant $\epsilon > 0$ such that the communication complexity of $(\epsilon, \epsilon)$-weak approximate Nash equilibrium in n-player binary-action games is at least $2^{\epsilon n}$.*

The proof follows similar lines to those in the proof of Theorem 2.1. Rather than two players whose actions correspond to $\Theta(n)$-long vectors, we have a player for each bit of (an encoding of) those vectors. We construct a game with $8m'$-players for $m' = \Theta(n)$ such that Alice holds the utility function of (the first) $3m'$ players, Bob holds the utilities of (the last) $5m'$ players, Alice's players utilities depend only on the $\alpha$s, Bob's utilities depend only on the $\beta$s, and every $(\epsilon^5/82, \epsilon^5/82)$-weak approximate Nash equilibrium corresponds to a $\delta$-fixed point of the function $f$ from Proposition 4.8.

**Players and Actions**

In section 4.5 we have used error correcting code to encode vertices that are deduced from $\mathbf{x}$ and $\mathbf{y}$. Here, since we consider *weak* approximate equilibria, we should add additional encodings for $I_v^A, I_v^A, v^B, w^B, \beta_v^B$ and $\beta_w^B$. Since we want to use the same number of players for each of the above objects, it will be convenient to encode them in the same space $\{0,1\}^{m'}$. We let the following be encoding functions of binary error correcting codes with constant (relative) distance:

- $E_I : \{0,1\}^3 \to \{0,1\}^{m'}$.

- $E_u : \{0,1\}^{2n+\log(n+1)} \to \{0,1\}^{m'}$.

- $E_\beta : \{0,1\}^{3\log M} \to \{0,1\}^{m'}$ (note that $3\log M = \Theta(n)$).

Let $E$ and $m$ denote encoding function and block length of the error correcting code from Section 4.4, i.e.:

- $E : \{0,1\}^{2n+\log(n+1)} \to \{0,1\}^m$.

For vectors $\mathbf{x}, \mathbf{y} \in [-1,2]^{4m}$, we use $(\frac{3}{\epsilon} - 1)$ bits to encode each continuous coordinate (up to precision $\epsilon$) in unary encoding. We choose $m'$ such that $m' = 4(\frac{3}{\epsilon} - 1)m$, so the encoding of each of $\mathbf{x}, \mathbf{y}$ also takes $m'$ bits. (For $E_\beta$, we must also have $m' > 3\log M$.) Here and henceforth, $\epsilon$ is a sufficiently small constant, satisfying $\epsilon = \Theta(\delta)$.

Instead of having a single player, Alice, with actions $(\mathbf{x}, I_v^A, I_w^A) \in \{-1, -1+\epsilon, \ldots, 2-\epsilon, 2\}^m \times \{0,1\}^3 \times \{0,1\}^3$ we replace her by $3m'$ players with binary actions. We have three types of Alice players:

- $\mathbf{x}$-type players. Player $\mathbf{x}_j^i$ chooses one of the actions $a_j^i \in \{j, j+\epsilon\}$ for every $i \in [4m]$ and $j \in \{-1, -1+\epsilon, \ldots, 2-2\epsilon, 2-\epsilon\}$. Note that the total number of $\mathbf{x}$-type players is $4m(\frac{3}{\epsilon} - 1) = m'$.

- $I^v$-type and $I^w$-type players. Player $I_i^v$ chooses a bit 0 or 1 for every $i \in [m']$. Similarly for $I^w$-type players.

In the communication problem, we assume that Alice knows the utilities of all the above players.

Instead of having a single player, Bob, with actions $(\mathbf{y}, v^B, w^B, \beta_v^B, \beta_w^B) \in \{-1, -1+\epsilon, \ldots, 2-\epsilon, 2\}^m \times \{0,1\}^{2n+\log(n+1)} \times \{0,1\}^{2n+\log(n+1)} \times [M]^3 \times [M]^3$ we replace him by $5m'$ players with binary actions. We have five types of players:

- $\mathbf{y}$-type players. Player $\mathbf{y}_j^i$ chooses one of the actions $b_j^i \in \{j, j+\epsilon\}$ for every $i \in [4m]$ and $j \in \{-1, -1+\epsilon, \ldots, 2-2\epsilon, 2-\epsilon\}$.

- $v$-type players. Player $v_i$ chooses a bit 0 or 1 for every $i \in [m']$. Similarly for $w$-type players.

- $\beta^v$-type players. Player $\beta_i^v$ chooses a bit 0 or 1 for every $i \in [m']$. Similarly for $\beta^w$-type players.

In the communication problem, we assume that Bob knows the utilities of all the above players.

**Utilities**

Before getting to the description of the utilities we define the notions of *realized number* and *realized point* by a set of players. For every $i \in [m]$, for simplicity of notations we add a dummy player $\mathbf{x}_2^i$ who has a single action $a_2^i = 2$. Given an action profile $a^i = (a_{-1}^i, a_{-1+\epsilon}^i, \ldots, a_2^i)$ of the players $\{\mathbf{x}_j^i\}_j$, the *realized number* $r(a^i) \in [-1, 2]$ is defined to be the minimal $j$ such that $a_j^i = j$. Note that $r(a_i)$ is well defined because the last player $\mathbf{x}_2^i$ plays 2. Given an action profile $a = (a_j^i)_{i,j}$ of all $\mathbf{x}$-type players we denote by $r(a) = (r(a_i))_i \in [-1,2]^m$ the realized point. Similarly we define the realized point of $\mathbf{y}$-type players.

The utilities are defined similarly to the two-player case with the following differences:

1. **x**-type/ **y**-type players' utilities are defined with respect to the *realized* points of the opponents. In addition, player that is responsible to the $i$-th coordinate of the point pays the distance from the $i$-th coordinate of the opponent's point/the $i$th coordinate of the $f$ operation of the opponent's point.

2. For all other types, the $i$-th player chooses the value of the $i$-th bit in the (alleged) codeword in $\{0,1\}^{m'}$.

Formally the payoffs are defined as follows:

- For **x**-type players, $U^{\mathbf{x}_j^i}\left(a_j^i; b^i\right) \triangleq -|a_j^i - r(b^i)|^2$, where we recall that player $\mathbf{x}_j^i$ is allowed to choose only $a_{i,j} = j$ or $a_{i,j} = j + \epsilon$, and $b^i$ is the profile of action played by players $\{\mathbf{y}_j^i\}_j$.

- For a $v$-type player $v_i$, we define $U^{v_i}(v_i; a) = 1$ iff he announces the bit $[E_u(D_v(r(a)))]_i$ (where the decoding function $D_v$ is defined in Section 4.5.1). Otherwise, $U^{v_i}(v_i; a) = 0$. Namely, the $i$-th player tries to guess the $i$-th coordinate of the encoded vector $E_u(v) \in \{0,1\}^{m'}$, were $v$ is computed using the decoding operation $D_v$ on the realized point $r(a) \in [-1, 2]^{4m}$. Similarly we define the utility of a $w$-type player.

- For a $\beta^v$-type player $\beta_i^v$, we define $U^{\beta_i^v}(\beta_i^v; a) = 1$ iff he announces the bit $[E_\beta(\beta_{D_v(r(a))}^S)]_i$. Namely, the $i$-th player tries to guess the $i$-th coordinate of the encoded vector $E_\beta(\beta_v^S)$, were $v$, as in the previous bullet, is computed using decoding. Similarly we define the utilities of $\beta^w$-type players.

- For a $I^v$-type player $I_i^v$, we define $U^{I_i^v}(I_i^v, \beta^v) = 1$ iff she announces the bit $\left[E_u\left(\alpha_{\overline{v}}^T([\overline{\beta}]_1), \alpha_{\overline{v}}^S([\overline{\beta}]_2), \alpha_{\overline{v}}^P([\overline{\beta}]_3)\right)\right]_i$ where $\overline{v}$ is the decoded vertex announced by $v$-type players and $\overline{\beta}$ is the decoded vector of indexes announced by $\beta^v$-type players. Similarly we define the utilities of $I^w$-type players.

- For **y**-type players, $U_{\mathbf{y}_j^i} = -|b_j^i - f_{\overline{I^v}, \overline{I^w}}(r(a))|^2$, where $\overline{I^v}$ and $\overline{I^w}$ are the decoding of the vertices announced by $I^v$-type and $I^w$-type players. We recall that the function $f_{I^v, I^w}$ is defined in Proposition 4.8.

### 4.6.1 Analysis of game

We analyse $(\overline{\epsilon}, \overline{\epsilon})$-weak approximate equilibria for $\overline{\epsilon} = \epsilon^5/82$. The analysis of the game follows the same sequence of Lemmas as the analysis in the two-player case (Section 4.5.2). The analogue of Lemma 4.9 is the following.

**Lemma 4.16.** *In every $(\overline{\epsilon}, \overline{\epsilon})$-weak approximate equilibrium $(\mathcal{A}, \mathcal{B})$, the realized point by the **x**-type players $r(a)$ satisfies*

$$\|r(a) - \mathsf{E}_{b \sim \mathcal{B}}[r(b)]\|_2^2 \leq \epsilon^2 \tag{18}$$

*with high probability[9] (the probability is over the mixed strategy of the **x**-type players).*

---

[9]Here and throughout this section, we use "with high probability" to mean with probability approaching 1 as $n$ grows (in fact, with an exponential dependence); in particular, the probability is approaching 1 faster than any polynomial in $\epsilon$.

*Proof.* We say that player $\mathbf{x}_j^i$'s action $j$ is *wrong* if $\mathsf{E}_{b^i \sim \mathcal{B}}[r(b^i)] \geq j + \epsilon$; similarly, we say that action $j + \epsilon$ is *wrong* $\mathsf{E}_{b^i \sim \mathcal{B}}[r(b^i)] \leq j$. Note that if for some coordinate $i$, *no* player $\mathbf{x}_j^i$ plays a wrong action, then the realized number $r_i(a_i)$ is $\epsilon$-close to $\mathsf{E}_{b^i \sim \mathcal{B}}[r(b^i)]$. We show that indeed in an $(\bar{\epsilon}, \bar{\epsilon})$-weak approximate equilibrium we will have many such coordinates $i$.

Recall that player $\mathbf{x}_j^i$'s utility when she plays $j$ is given by

$$u(j) \triangleq \mathsf{E}_{b^i \sim \mathcal{B}}\left[U^{\mathbf{x}_j^i}(j; b^i)\right] = \mathsf{E}_{b^i \sim \mathcal{B}}[-|j - r(b^i)|^2] = -|j - \mathsf{E}_{b^i \sim \mathcal{B}}[r(b^i)]|^2 - \mathrm{Var}_{b^i \sim \mathcal{B}}[r(b^i)].$$

Similarly, when she plays $j + \epsilon$ her utility is given by

$$u(j + \epsilon) \triangleq \mathsf{E}_{b^i \sim \mathcal{B}}\left[U^{\mathbf{x}_j^i}(j + \epsilon; b^i)\right] = \mathsf{E}_{b^i \sim \mathcal{B}}[-|j + \epsilon - r(b^i)|^2] = -|j + \epsilon - \mathsf{E}_{b^i \sim \mathcal{B}}[r(b^i)]|^2 - \mathrm{Var}_{b^i \sim \mathcal{B}}[r(b^i)].$$

When $j$ is wrong (i.e. $\mathsf{E}_{b^i \sim \mathcal{B}}[r(b^i)] \geq j + \epsilon$) the difference in the utilities $u(j + \epsilon) - u(j)$ is given by

$$\begin{aligned} u(j + \epsilon) - u(j) &= -(\mathsf{E}_{b^i \sim \mathcal{B}}[r(b^i)] - j - \epsilon)^2 + (\mathsf{E}_{b^i \sim \mathcal{B}}[r(b^i)] - j)^2 \\ &= (2\mathsf{E}_{b^i \sim \mathcal{B}}[r(b^i)] - 2j - \epsilon)\epsilon \geq \epsilon^2 \end{aligned}$$

For $j + \epsilon$ is wrong ($\mathsf{E}_{b^i \sim \mathcal{B}}[r(b^i)] \leq j$) the difference in the utilities $u(j) - u(j + \epsilon)$ is given by

$$\begin{aligned} u(j) - u(j + \epsilon) &= -(j - \mathsf{E}_{b^i \sim \mathcal{B}}[r(b^i)])^2 + (j + \epsilon - \mathsf{E}_{b^i \sim \mathcal{B}}[r(b^i)])^2 \\ &= (2j - 2\mathsf{E}_{b^i \sim \mathcal{B}}[r(b^i)] + \epsilon)\epsilon \geq \epsilon^2 \end{aligned}$$

Therefore, player $\mathbf{x}_j^i$ can always increase her payoff by at least $\epsilon^2$ by deviating from a wrong action. Note that if player $\mathbf{x}_j^i$ is $\bar{\epsilon}$-best replying, she assigns a probability of at most $\bar{\epsilon}/\epsilon^2$ to a wrong action. In addition, the fraction of $\mathbf{x}$-type players that are not $\bar{\epsilon}$-best replying is at most $8\bar{\epsilon}$ (because we have 8 types of players of equal cardinality). Therefore, in the expected fraction of $\mathbf{x}$-type players playing a wrong is at most $8\bar{\epsilon} + 2\bar{\epsilon}/\epsilon^2 < 2.5\bar{\epsilon}/\epsilon^2$. Therefore, with high probability over $\mathbf{x}$-type players mixed strategies, at most a $3\bar{\epsilon}/\epsilon^2$-fraction play a wrong action (e.g. by Chernoff bound). Therefore the fraction of coordinates $i \in [4m]$ where at least one player $\mathbf{x}_j^i$ plays a wrong action is at most $9\bar{\epsilon}/\epsilon^3$ (because we have $3/\epsilon$ players in each coordinate). So in $(1 - 9\bar{\epsilon}/\epsilon^3)$ fraction of coordinates we have $|r_i(a_i) - \mathsf{E}_{b^i \sim \mathcal{B}}[r(b^i)]| \leq \epsilon$, which implies

$$\|r(a) - \mathsf{E}_{b \sim \mathcal{B}}[r(b)]\|_2^2 = \frac{1}{4m} \sum_i |r(a_i) - \mathsf{E}_{b \sim \mathcal{B}}[r(b^i)]|^2 \leq (1 - \frac{9\bar{\epsilon}}{\epsilon^3})\epsilon^2 + \frac{9\bar{\epsilon}}{\epsilon^3} 3^2 < \frac{82\bar{\epsilon}}{\epsilon^3} = \epsilon^2$$

$\square$

The analogue of Lemma 4.10 is the following.

**Lemma 4.17.** *In every $(\bar{\epsilon}, \bar{\epsilon})$-weak approximate equilibrium $(\mathcal{A}, \mathcal{B})$, if the first $m$-tuple of coordinates of $\mathsf{E}_{b \sim \mathcal{B}}[r(b)]$ is $6h$-close to the binary encoding $E(v)$ of a vertex $v$, then*

1. *The decoding of the action profile of the $v$-type players is $v$ with probability $1 - o(\epsilon)$.*

2. *The decoding of the action profile of the $\beta^v$-type players is $(\beta_v^T, \beta_v^S, \beta_v^P)$ with probability $1 - o(\epsilon)$.*

*Proof.* Whenever (18) holds, $D_v(r(a)) = v$. In particular, for each $i \in [m']$, $[E_u(D_v(r(a)))]_i = [E_u(v)]_i$ with high probability. Therefore, by playing the action $[E_u(v)]_i$ player $v_i$ has expected utility of $1 - o(1)$ whereas by playing the action $1 - [E_u(v)]_i$ his expected utility is $o(1)$.

Every player that is $\bar{\epsilon}$-best replying, assigns probability of at least $1 - O(\bar{\epsilon})$ to the correct bit. In addition, we have at most $8\bar{\epsilon}$ fraction of $v$-type players who are not $\bar{\epsilon}$-best replying (because we have 8 types of players of equal cardinality). Therefore the expected fraction of $v$-type players who play the wrong bit is $O(\bar{\epsilon})$. By Chernoff bound, it also holds that with high probability at most an $O(\bar{\epsilon})$-fraction of $v$-type players play the wrong bit. Whenever this is the case, $v$ is indeed decoded correctly.

Similarly we prove the second claim in the lemma for $\beta^v$-type players. □

In a similar way we can show that analogues of Lemmas 4.11, 4.12, 4.13, and 4.14 hold for the $n$-player game. In particular,

**Lemma 4.18.** *In every $(\bar{\epsilon}, \bar{\epsilon})$-weak approximate equilibrium $(\mathcal{A}, \mathcal{B})$, $f_{I_v^A, I_w^A}(\mathbf{x}) = f(\mathbf{x})$ with high probability.*

Now we get to the analogue of the last Corollary 4.15.

**Corollary 4.19.** *In every $(\bar{\epsilon}, \bar{\epsilon})$-weak approximate equilibrium $(\mathcal{A}, \mathcal{B})$, the expectation of the realized point $\mathsf{E}_{a \sim \mathcal{A}}[r(a)]$ is a $\delta$-approximate equilibrium of $f$; i.e.,*

$$\|\mathsf{E}_{a \sim \mathcal{A}}[r(a)] - f(\mathsf{E}_{a \sim \mathcal{A}}[r(a)])\|_2 \leq \delta.$$

*Proof.* The proof is similar to the proof of Corollary 4.15. We recall that in Lemma 4.16 we have proved that

$$\|r(a) - \mathsf{E}_{b \sim \mathcal{B}}[r(b)]\|_2^2 \leq \epsilon^2 \tag{19}$$

with high probability. This, in particular, implies that $r(a)$ is, with high probability, close to its expectation:

$$\|r(a) - \mathsf{E}_{a' \sim \mathcal{A}}[r(a')]\|_2^2 \leq 2\|r(a) - \mathsf{E}_{b \sim \mathcal{B}}[r(b)]\|_2^2 + 2\|\mathsf{E}_{a \sim \mathcal{A}}[r(a)] - \mathsf{E}_{b \sim \mathcal{B}}[r(b)]\|_2^2 \quad \text{(Triangle ineq.)}$$

$$\leq 2\|r(a) - \mathsf{E}_{b \sim \mathcal{B}}[r(b)]\|_2^2 + 2\mathsf{E}_{a' \sim \mathcal{A}}\left[\|r(a') - \mathsf{E}_{b \sim \mathcal{B}}[r(b)]\|_2^2\right] \quad \text{(Convexity)}$$

$$= O(\epsilon^2), \quad \text{(Lemma 4.16)}$$
$$\tag{20}$$

with high probability.

Using the $O(1)$-Lipschitzness of $f$ we deduce that

$$\|f(r(a)) - f(\mathsf{E}_{a' \sim \mathcal{A}}[r(a')])\|_2^2 = O(\epsilon^2) \tag{21}$$

with high probability.

Using similar arguments to those of Lemma 4.16 we can show that

$$\|r(b) - \mathsf{E}_{a' \sim \mathcal{A}}[f_{\overline{I^v}, \overline{I^w}}(r(a'))]\|_2^2 = O(\epsilon^2) \tag{22}$$

with high probability, where we recall that $\overline{I^v}, \overline{I^w}$ denote the decoded line information of the action profile played by the $I^v, I^w$-types players. By an analogous argument to (20),

$$\|r(b) - \mathsf{E}_{b' \sim \mathcal{B}}[r(b')]\|_2^2 = O(\epsilon^2) \tag{23}$$

with high probability.

By Lemma 4.18,

$$\left\|\mathsf{E}_{a'\sim\mathcal{A}}[f_{\overline{I^v},\overline{I^w}}(r(a'))] - \mathsf{E}_{a'\sim\mathcal{A}}[f(r(a'))]\right\|_2^2 = O(\epsilon^2). \tag{24}$$

By Equations (20),(19),(23),(22),(24),(21) (applied exactly in this order) and the triangle inequality we get

$$\left\|\mathsf{E}_{a\sim\mathcal{A}}[r(a')] - f(\mathsf{E}_{a\sim\mathcal{A}}[r(a')])\right\|_2^2 = O(\epsilon^2) < \delta^2. \tag{25}$$

$\square$

*Proof of Theorem 2.2.* Any communication protocol that solves the $(\epsilon^5/82, \epsilon^5/82)$-weak approximate Nash equilibrium problem in $\Theta(n)$-player games with binary actions induces a communication protocol for the problem SIM-EOAL: Alice constructs the utilities of her players using her private information of the $\alpha$s, Bob constructs his utility using the $\beta$s. They implement the communication protocol to find an $(\epsilon^5/82, \epsilon^5/82)$-weak approximate Nash equilibrium, and then both of them know $\mathsf{E}_{a\sim\mathcal{A}}[r(a)]$ which is a $\delta$-approximate fixed point of $f$ (by Corollary 4.19). Finally, they round and decode the approximate fixed point to get an end or start of a line.

Using Corollary 4.7 we deduce that the communication complexity of $(\epsilon^5/82, \epsilon^5/82)$-weak approximate Nash equilibrium problem in $\Theta(n)$-player games with binary actions is at least $2^{\Omega(n)}$. $\square$

## 5 An Open Problem: Correlated Equilibria in 2-Player Games

As mentioned in Section 1.3, it is known that for $n$-player, $O(1)$-action games, even *exact* correlated equilibrium can be found with $\mathsf{poly}(n)$ deterministic communication complexity (see [HM10, PR08, JLB15]).

For approximate correlated equilibrium in two-player $N \times N$ games, to the best of our knowledge, no non-trivial results are known (neither positive nor negative). Does a $\mathsf{polylog}(N)$ communication protocol for approximate correlated equilibrium exist? Is there a $\mathsf{poly}(N)$ communication lower bound?

## Acknowledgements

We are very grateful for Mika Goos, who have pointed out pointed that the WAPP Simulation Theorem can be applied in our case, which allowed us to generalize the hardness result from deterministic to randomized communication complexity. We also thank Noam Nisan for inspiring discussions and suggestions.

# A  Proof of Proposition 4.8

**Proposition** (Proposition 4.8, restated)**.** *There exist constants[10] $\delta, h > 0$ such that given a line $I = (T, S, P)$ over $G$ there exists a function $f = f(I) : [-1,2]^{4m} \to [-1,2]^{4m}$ and a class of functions $\{f_{I_1, I_2} : [-1,2]^{4m} \to [-1,2]^{4m}\}_{I_i \in \{0,1\}^3}$ which do not depend on $I$ with the following properties:*

1. *$\|f(\mathbf{x}) - \mathbf{x}\|_2 = \Omega(\delta)$ for every $\mathbf{x}$ that in not $2\sqrt{h}$-close to the embedded edge of any non-trivial end or start of a line (i.e., the embedding of the edge $(P(v^*), v^*)$ such that $T(v^*) = 1$ but $T(S(v^*)) = 0$; or the edge $(v^*, S(v^*))$ for $v^*$ such that $T(P(v^*)) = 0$, $T(v^*) = 1$, and $v^* \neq \mathbf{0}_{2n+1}$).*

2. *$f$ is $O(1)$-Lipschitz in $\|\cdot\|_2$ norm.*

3. *$f$ is local in the sense that there exists a class of (sixty four) functions $\{f_{I_1, I_2} : [-1,2]^{4m} \to [-1,2]^{4m}\}_{I_i \in \{0,1\}^3}$ which do not depend on $I$, and $f$ can be defined as an interpolation between these functions such that:*

---

[10]In the restated Proposition we have changed the values of the constant $\delta$ and $h$. The constant $\delta$ in the original proposition is replaced by $c\delta$, for sufficiently large constant $c$, in the present restatement. The constant $h$ in the original proposition is replaced by $2\sqrt{h}$ in the present restatement. The reason for this change in constants will become clear in the proof, where the current $\delta$ and $h$ will have a natural meaning.

(a) If the first $m$-tuple of coordinates of $\mathbf{x}$ is $12\sqrt{h}$-close to the encoded vertex $E(v)$, but the second $m$-tuple of coordinates of $\mathbf{x}$ is $12\sqrt{h}$-far from any encoded vertex $E(w)$ then $f_{I(v),I_2}(\mathbf{x}) = f(\mathbf{x})$ for every $I_2 \in \{0,1\}^3$.

(b) If the second $m$-tuple of coordinates of $\mathbf{x}$ is $12\sqrt{h}$-close to the encoded vertex $E(w)$, but the first $m$-tuple of coordinates of $\mathbf{x}$ is $12\sqrt{h}$-far from any encoded vertex $E(v)$ then $f_{I_1,I(w)}(\mathbf{x}) = f(\mathbf{x})$ for every $I_1 \in \{0,1\}^3$.

(c) If the first $m$-tuple of coordinates of $\mathbf{x}$ is $12\sqrt{h}$-close to the encoded vertex $E(v)$, and the second $m$-tuple of coordinates of $\mathbf{x}$ is $12\sqrt{h}$-close to the encoded vertex $E(w)$ then $f_{I(v),I(w)}(\mathbf{x}) = f(\mathbf{x})$.

(d) If none of the above conditions are satisfied, then $f_{I_1,I_2}(\mathbf{x}) = f(\mathbf{x})$ for every $I_1, I_2 \in \{0,1\}^3$.

In Section A.1 we recall the notations of the embedding that was introduced in Section 4.4. In Section A.2 we construct the function $f$. In Section A.3 we construct the class $\{f_{I_1,I_2}\}_{I_i \in \{0,1\}^3}$, using the locality of $f$.

## A.1 Discrete embedding of a graph in the Euclidean space

For convenience, we recall here the notations of our discrete embedding of $G$ in $[-1,2]^{4m}$ from Section 4.4.

The vertex $v$ is embedded to the point $(E(v), E(v), \mathbf{0}_m, \mathbf{0}_m) \in [-1,2]^{4m}$, which is called *the embedded vertex*.

For every edge $(v,w)$ in $G$, we define five vertices:

$$\mathbf{x}^1(v,w) \triangleq (E(v), E(v), \mathbf{0}_m, \mathbf{0}_m)$$
$$\mathbf{x}^2(v,w) \triangleq (E(v), E(v), \mathbf{1}_m, \mathbf{0}_m)$$
$$\mathbf{x}^3(v,w) \triangleq (E(v), E(w), \mathbf{1}_m, \mathbf{0}_m)$$
$$\mathbf{x}^4(v,w) \triangleq (E(v), E(w), \mathbf{0}_m, \mathbf{0}_m)$$
$$\mathbf{x}^5(v,w) \triangleq (E(w), E(w), \mathbf{0}_m, \mathbf{0}_m).$$

The vertices $\mathbf{x}^i(v,w)$ are called *Brouwer vertices*. Note that $\mathbf{x}^1(v,w)$ is the embedded vertex $v$, $\mathbf{x}^5(v,w)$ is the embedded vertex $w$. The line that connects the points $\mathbf{x}^i(v,w)$ and $\mathbf{x}^{i+1}(v,w)$ is called a *Brouwer line segment*. The union of these four Brouwer line segments is called the *embedded edge* $(v,w)$.

## A.2 The function $f$

We set $h$ to be a sufficiently small constant such that the $\sqrt{h}$ neighbourhood of any two Brouwer vertices will not intersect and such that the $3h$ neighbourhood of any two Brouwer line segments will not intersect- unless they share the same common Brouwer vertex. We take $\delta$ to be a constant arbitrarily smaller than $h$ ($\delta = h^3$ suffices). We define a displacement function $g : [-1,2]^{4m} \to [-\delta,\delta]^{4m}$ and $f(\mathbf{x}) \triangleq \mathbf{x} + g(\mathbf{x})$. In order that Properties (1)-(3) of Proposition 4.8 will be satisfied, we should define $g$ such that:

1. $\|g(\mathbf{x})\|_2 = \Omega(\delta)$ for every $x$ that is not $2\sqrt{h}$-close to the Brouwer line segments of any non-trivial end or starting of a line.

2. $g$ is $O(1)$-Lipschitz.

3. $g$ is defined "locally", which will allow as to generate the class of functions $\{f_{I_1,I_2}\}$.

We think of the $4m$ coordinates as partitioned into four parts: the first $m$-tuple of coordinates represent the current vertex in the line; the second $m$-tuple represent the next vertex in the line; we think of the third $m$-tuple as all being equal to a single bit that monitors helps altering between computing the next vertex, and copying from the second to first $m$-tuple. Finally, the last $m$ coordinates represent a special default direction in which the displacement points when far from all Brouwer line segments (similarly to the single special coordinate in [HPV89]).

We consider a path starting at $(\mathbf{0}_{3m}, 2 \cdot \mathbf{1}_m)$, i.e. the concatenation of 0 on the first $3m$ coordinates, and 2 on the last $m$ coordinates. The path first goes to $(\mathbf{0}_{4m})$ (in a straight line), and thereafter the last $m$ coordinates remain constantly 0 (note that every Brouwer vertex has $\mathbf{0}_m$ in its last $m$-tuple). The first $3m$ coordinates follow the line according to the embedding in Section 4.4. This path corresponds to the line starting at $\mathbf{0}_{2n+1}$; for any additional line starting at vertex $u$, we have another path starting at $(E(u), E(u), \mathbf{0}_{2m})$.

We say that a point $\mathbf{x}$ is in the *picture* if $\frac{1}{m}\sum_{i=3m+1}^{4m} x_i < 1/2$. We construct $g$ separately inside and outside the picture (and make sure that the construction agrees on the hyperplane $\frac{1}{m}\sum_{i=3m+1}^{4m} x_i = 1/2$).

**Truncation** In order for $g(\cdot)$ to be a displacement function, we must ensure that it never sends any points outside the hypercube, i.e. $\forall \mathbf{x} \in [-1,2]^{4m}$, we require that also $\mathbf{x} + g(\mathbf{x}) \in [-1,2]^{4m}$. Below, it is convenient to first define an *untruncated* displacement function $\hat{g} : [-1,2]^{4m} \to [-\delta, \delta]^{4m}$ which is not restricted by the above condition. We then truncate each coordinate to fit in $[-1,2]$: $[g(\mathbf{x})]_i = \max\{-1, \min\{2, x_i + [\hat{g}(\mathbf{x})]_i\}\} - x_i$. It is clear that if $\hat{g}(\cdot)$ is $(M-1)$-Lipschitz, then $g(\cdot)$ is $M$-Lipschitz. It is, however, important to make sure that the magnitude of the displacement is not compromised. Typically, some of the coordinates may need to be truncated, but we design the displacement so that most coordinates, say 99%, are not truncated. If $\hat{g}(\mathbf{x})$ has a non-negligible component in at least 5% of the coordinates, then in total $g(\mathbf{x})$ maintains a non-negligible magnitude.

### A.2.1 Inside the picture

The line $\mathbf{0} = v_0, v_1, \ldots, v^*$ in $G$ is embedded to a path in $[-1,2]^{4m}$ that goes in straight lines through the following sequence of Brouwer vertices:

$$(\mathbf{0}_{3m}, 2 \cdot \mathbf{1}_m), (\mathbf{0}_{4m}) = \mathbf{x}^1(v_0, v_1), \mathbf{x}^2(v_0, v_1), \ldots, \mathbf{x}^5(v_0, v_1) = \mathbf{x}^1(v_1, v_2), \mathbf{x}^2(v_1, v_2), \ldots, \mathbf{x}^5(P(v^*), v^*).$$

Similarly, if $I$ contains another line $u, \ldots, w$, it is embedded as a path through:

$$(E(u), E(u), \mathbf{0}_{2m}) = \mathbf{x}^1(u, S(u)), \ldots, \mathbf{x}^5(P(w), w) = (E(w), E(w), \mathbf{0}_{2m}).$$

Now we *cut the corners* of this path as follows: For two consecutive Brouwer vertices $\mathbf{s}, \mathbf{y}$ in the embedded path we let $\mathbf{z}^1_{(\mathbf{s} \to \mathbf{y})}$ be the point in the Brouwer line segment $[\mathbf{s}, \mathbf{y}]$ that is exactly $\sqrt{h}$-far from $\mathbf{s}$. Similarly, $\mathbf{z}^2_{(\mathbf{s} \to \mathbf{y})}$ is the point in $[\mathbf{s}, \mathbf{y}]$ that is exactly $\sqrt{h}$-far from $\mathbf{y}$. For three consecutive Brouwer vertices $\mathbf{s} \to \mathbf{y} \to \mathbf{t}$, the path after "cutting the corners" goes in straight lines through

$$\ldots, \mathbf{z}^1_{(\mathbf{s} \to \mathbf{y})}, \mathbf{z}^2_{(\mathbf{s} \to \mathbf{y})}, \mathbf{z}^1_{(\mathbf{y} \to \mathbf{t})}, \mathbf{z}^2_{(\mathbf{y} \to \mathbf{t})}, \ldots$$

instead of going through $\mathbf{s} \to \mathbf{y} \to \mathbf{t}$.

First, for all points inside the picture that are $3h$-far from the embedded path after cutting the corners we use the same *default displacement* which points in the positive special direction: $\hat{g}(\mathbf{x}) = (\mathbf{0}_{3m}, \delta \cdot \mathbf{1}_m)$. Because $\mathbf{x}$ is inside the picture, the truncated displacement $g(\mathbf{x})$ is close to $\hat{g}(\mathbf{x})$, and therefore satisfies $\|g(\mathbf{x})\|_2 = \Omega(\delta)$.

Now we define the displacement $3h$-close to the embedded path in two regions:

1. For points that are $3h$-close to a segment of the form $[\mathbf{z}^1_{(\mathbf{s} \to \mathbf{y})}, \mathbf{z}^2_{(\mathbf{s} \to \mathbf{y})}]$ but (approximately[11]) $\sqrt{h}$-far from both Brouwer vertices $\mathbf{s}, \mathbf{y}$.

2. For the remaining points, those that are $3h$-close to a segment of the form $[\mathbf{z}^2_{(\mathbf{s} \to \mathbf{y})}, \mathbf{z}^1_{(\mathbf{y} \to \mathbf{t})}]$ and (approximately[11]) $\sqrt{h}$-close to the Brouwer vertex $\mathbf{y}$.

We make sure that the definitions agree on the interface between the two regions, as well as on the interface with the points that receive the default displacement.

### A.2.2 Close to the path but far from a Brouwer vertex

On the Brouwer line segment, the displacement points in the direction of the path; at distance $h$ from the Brouwer line segment, the displacement points in towards the Brouwer line segment; at distance $2h$ from the Brouwer line segment, the displacement points against the direction of the path; at distance $3h$, the displacement points in the default direction.

Formally, let $\sigma_{(\mathbf{s} \to \mathbf{t})}(\mathbf{x})$ denote the magnitude of the component of $\mathbf{x} - \mathbf{s}$ in the direction of line $(\mathbf{s} \to \mathbf{t})$,

$$\sigma_{(\mathbf{s} \to \mathbf{t})}(\mathbf{x}) \triangleq \frac{(\mathbf{t} - \mathbf{s})}{\|\mathbf{s} - \mathbf{t}\|_2} \cdot (\mathbf{x} - \mathbf{s}),$$

where $\cdot$ denotes the (in-expectation) dot product. Let $\mathbf{z} = \mathbf{z}(\mathbf{x})$ be the point nearest to $\mathbf{x}$ on the Brouwer line segment; notice that $\mathbf{z}$ satisfies

$$\mathbf{z} = \sigma_{(\mathbf{s} \to \mathbf{t})}(\mathbf{x}) \mathbf{t} + (1 - \sigma_{(\mathbf{s} \to \mathbf{t})}(\mathbf{x})) \mathbf{s}.$$

For points near the Brouwer line segment ($\|\mathbf{x} - \mathbf{z}\|_2 \le 3h$), but far from its endpoints ($\sigma_{(\mathbf{s} \to \mathbf{t})}(\mathbf{x}) \in [\sqrt{h}, 1 - \sqrt{h}]$), we define the displacement:

$$\hat{g}(\mathbf{x}) \triangleq \begin{cases} \delta \frac{(\mathbf{t}-\mathbf{s})}{\|\mathbf{t}-\mathbf{s}\|_2} & \|\mathbf{x} - \mathbf{z}\|_2 = 0 \\ \delta \frac{(\mathbf{z}-\mathbf{x})}{h} & \|\mathbf{x} - \mathbf{z}\|_2 = h \\ \delta \frac{(\mathbf{s}-\mathbf{t})}{\|\mathbf{t}-\mathbf{s}\|_2} & \|\mathbf{x} - \mathbf{z}\|_2 = 2h \\ \delta (\mathbf{0}_{3m}, \mathbf{1}_m) & \|\mathbf{x} - \mathbf{z}\|_2 = 3h \end{cases} \tag{26}$$

At intermediate distances from the Brouwer line segment, we interpolate: at distance $\|\mathbf{x} - \mathbf{z}\|_2 = \frac{1}{3}h$, for example, we have $\hat{g}(\mathbf{x}) = \frac{2}{3}\delta \frac{(\mathbf{t}-\mathbf{s})}{\|\mathbf{t}-\mathbf{s}\|_2} + \frac{1}{3}\delta \frac{(\mathbf{z}-\mathbf{x})}{h}$. Notice that every two of $(\mathbf{t} - \mathbf{s})$, $(\mathbf{z} - \mathbf{x})$, and $(\mathbf{0}_{3m}, \mathbf{1}_m)$ are orthogonal, so the interpolation does not lead to cancellation. Also, every point $\mathbf{z}$ on the Brouwer line segment is $\Omega(1)$-far in every coordinate from $\{-1, 2\}$, so the truncated displacement $g(\mathbf{x})$ still satisfies $\|g(\mathbf{x})\|_2 = \Omega(\delta)$. For each case in (26), $\hat{g}(\cdot)$ is either constant, or (in the case of $\|\mathbf{x} - \mathbf{z}\|_2 = h$) $O(\delta/h)$-Lipschitz ($\frac{(\mathbf{z}-\mathbf{x})}{h}$

---

[11] It will be more convenient to set the threshold of points $\mathbf{x}$ that are "far"/"close" from/to a Brouwer vertex using the expression $\sigma_{(\mathbf{s} \to \mathbf{y})}(\mathbf{x})$ that is defined below. $\sigma_{(\mathbf{s} \to \mathbf{y})}(x)$ is closely related to the distance of $\mathbf{x}$ from the points $\mathbf{s}, y$ but is not precisely the distance.

is $O(1/h)$-Lipschitz because two "antipodal" points at distance $2h$ have opposite directions, both pointing parallel to the Brouwer line segment); by choice of $\delta \ll h$, it follows that $\hat{g}(\cdot)$ is in particular $O(1)$-Lipschitz. Furthermore, notice that $\|\mathbf{x} - \mathbf{z}\|_2$ is 1-Lipschitz, so after interpolating for intermediate distances, $\hat{g}(\cdot)$ continues to be $O(1)$-Lipschitz. Notice also that at distance $3h$ the displacement defined in (26) agrees with the displacements for points far from every Brouwer line segment, so Lipschitz continuity is preserved.

### A.2.3 Close to the path and a Brouwer vertex

Let $L_\mathbf{y}$ be the line that connects the points $\mathbf{z}_{(\mathbf{s} \to \mathbf{y})}$ and $\mathbf{z}_{(\mathbf{y} \to \mathbf{t})}$. Given $\mathbf{x}$, we let $\mathbf{z}$ be the closest point to $\mathbf{x}$ on $L_\mathbf{y}$.

Our goal is to interpolate between the line displacement for $(\mathbf{s} \to \mathbf{y})$ (which is defined up to $\sigma_{(\mathbf{s} \to \mathbf{y})}(\mathbf{x}) = 1 - \sqrt{h}$), and the line displacement for $(\mathbf{y} \to \mathbf{t})$ (which begins at $\sigma_{(\mathbf{y} \to \mathbf{t})}(\mathbf{x}) = \sqrt{h}$). Let $\Delta_{(\mathbf{s} \to \mathbf{y})}(\mathbf{x}) \triangleq \sigma_{(\mathbf{s} \to \mathbf{y})}(\mathbf{x}) - (1 - \sqrt{h})$, and $\Delta_{(\mathbf{y} \to \mathbf{t})}(\mathbf{x}) \triangleq \sqrt{h} - \sigma_{(\mathbf{y} \to \mathbf{t})}(\mathbf{x})$. We set our interpolation parameter $\tau = \tau(\mathbf{x}) \triangleq \frac{\Delta_{(\mathbf{y} \to \mathbf{t})}(\mathbf{x})}{\Delta_{(\mathbf{y} \to \mathbf{t})}(\mathbf{x}) + \Delta_{(\mathbf{s} \to \mathbf{y})}(\mathbf{x})}$, and set

$$\mathbf{z} \triangleq \tau \mathbf{z}_{(\mathbf{s} \to \mathbf{y})} + (1 - \tau) \mathbf{z}_{(\mathbf{y} \to \mathbf{t})}. \tag{27}$$

For points $\mathbf{x}$ near $\mathbf{y}$ such that $\Delta_{(\mathbf{s} \to \mathbf{y})}(\mathbf{x}), \Delta_{(\mathbf{y} \to \mathbf{t})}(\mathbf{x}) \geq 0$, we can now define the displacement analogously to (26):

$$\hat{g}(\mathbf{x}) \triangleq \begin{cases} \delta \cdot \left[\tau \frac{(\mathbf{y}\text{-}\mathbf{s})}{\|\mathbf{y}\text{-}\mathbf{s}\|_2} + (1-\tau) \frac{(\mathbf{t}\text{-}\mathbf{y})}{\|\mathbf{t}\text{-}\mathbf{y}\|_2}\right] & \|\mathbf{x} - \mathbf{z}\|_2 = 0 \\ \delta \frac{(\mathbf{z}\text{-}\mathbf{x})}{h} & \|\mathbf{x} - \mathbf{z}\|_2 = h \\ \delta \cdot \left[\tau \frac{(\mathbf{s}\text{-}\mathbf{y})}{\|\mathbf{y}\text{-}\mathbf{s}\|_2} + (1-\tau) \frac{(\mathbf{y}\text{-}\mathbf{t})}{\|\mathbf{t}\text{-}\mathbf{y}\|_2}\right] & \|\mathbf{x} - \mathbf{z}\|_2 = 2h \\ \delta(\mathbf{0}_{3m}, \mathbf{1}_m) & \|\mathbf{x} - \mathbf{z}\|_2 \geq 3h \end{cases}. \tag{28}$$

At intermediate distances, interpolate according to $\|\mathbf{x} - \mathbf{z}\|_2$. Notice that for each fixed choice of $\tau \in [0, 1]$ (and $\mathbf{z}$), $\hat{g}$ is $O(\delta/h) = O(1)$-Lipschitz. Furthermore, $\Delta_{(\mathbf{s} \to \mathbf{y})}$ and $\Delta_{(\mathbf{y} \to \mathbf{t})}$ are 1-Lipschitz in $\mathbf{x}$. For any $\mathbf{z} \in L_\mathbf{y}$, $\Delta_{(\mathbf{y} \to \mathbf{t})}(\mathbf{z}) + \Delta_{(\mathbf{s} \to \mathbf{y})}(\mathbf{z}) = \sqrt{h}$. For general $\mathbf{x}$, we have

$$\Delta_{(\mathbf{y} \to \mathbf{t})}(\mathbf{x}) + \Delta_{(\mathbf{s} \to \mathbf{y})}(\mathbf{x}) \geq \Delta_{(\mathbf{y} \to \mathbf{t})}(\mathbf{z}) + \Delta_{(\mathbf{s} \to \mathbf{y})}(\mathbf{z}) - 2\|\mathbf{x} - \mathbf{z}\|_2 = \sqrt{h} - 2\|\mathbf{x} - \mathbf{z}\|_2; \tag{29}$$

so $\tau$ is $O(1/\sqrt{h})$-Lipschitz whenever $\|\mathbf{x} - \mathbf{z}\|_2 < 3h$, and otherwise has no effect on $\hat{g}(\mathbf{x})$. We conclude that $\hat{g}$ is $O(1)$-Lipschitz when interpolating across different values of $\tau$. At the interface with (26) $\tau$ is 1 (0 near $\mathbf{z}_{(\mathbf{y} \to \mathbf{t})}$), so (26) and (28) are equal. Therefore $\hat{g}$ is $O(1)$-Lipschitz on all of $[-1, 2]^{4m}$.

To lower bound the magnitude of the displacement, we argue that $(\mathbf{z} - \mathbf{x})$ is orthogonal to $\left[\tau \frac{(\mathbf{y}\text{-}\mathbf{s})}{\|\mathbf{y}\text{-}\mathbf{s}\|_2} + (1-\tau) \frac{(\mathbf{t}\text{-}\mathbf{y})}{\|\mathbf{t}\text{-}\mathbf{y}\|_2}\right]$. First, observe that we can restrict our attention to the component of $(\mathbf{z} - \mathbf{x})$ that belongs to the plane defined by $\mathbf{s}, \mathbf{y}, \mathbf{t}$ (in which $\mathbf{z}$ also lies). Let $P_{\mathbf{s}, \mathbf{y}, \mathbf{t}}(\mathbf{x})$ denote the projection of $\mathbf{x}$ to this plain. We can write points in this plane in terms of their $\Delta(\cdot) \triangleq (\Delta_{(\mathbf{s} \to \mathbf{y})}(\cdot), \Delta_{(\mathbf{y} \to \mathbf{t})}(\cdot))$ values. (Recall that $(\mathbf{s} \to \mathbf{y})$ and $(\mathbf{y} \to \mathbf{t})$ are orthogonal.)

First, observe that $\Delta(\mathbf{z}_{(\mathbf{s} \to \mathbf{y})}) = (0, \sqrt{h})$, $\Delta(\mathbf{z}_{(\mathbf{y} \to \mathbf{t})}) = (\sqrt{h}, 0)$ and $\Delta(\mathbf{y}) = (\sqrt{h}, \sqrt{h})$. Notice also that

$$\left[\tau \frac{(\mathbf{y} - \mathbf{s})}{\|\mathbf{y} - \mathbf{s}\|_2} + (1 - \tau) \frac{(\mathbf{t} - \mathbf{y})}{\|\mathbf{t} - \mathbf{y}\|_2}\right] = \left[\tau \frac{(\mathbf{y} - \mathbf{z}_{(\mathbf{s} \to \mathbf{y})})}{\sqrt{h}} + (1 - \tau) \frac{(\mathbf{z}_{(\mathbf{y} \to \mathbf{t})} - \mathbf{y})}{\sqrt{h}}\right].$$

Putting those together, we have that

$$\Delta\left(\left[\tau\frac{\mathbf{y}}{\|\mathbf{y}-\mathbf{s}\|_2}+(1-\tau)\frac{\mathbf{t}}{\|\mathbf{t}-\mathbf{y}\|_2}\right]\right)-\Delta\left(\left[\tau\frac{\mathbf{s}}{\|\mathbf{y}-\mathbf{s}\|_2}+(1-\tau)\frac{\mathbf{y}}{\|\mathbf{t}-\mathbf{y}\|_2}\right]\right)=(\tau,1-\tau). \quad (30)$$

For $\mathbf{z}$, we have

$$\Delta(\mathbf{z}) = \tau\Delta\left(\mathbf{z}_{(\mathbf{s}\to\mathbf{y})}\right)+(1-\tau)\Delta\left(\mathbf{z}_{(\mathbf{y}\to\mathbf{t})}\right) = \sqrt{h}\left(1-\tau,\tau\right).$$

Finally, for $P_{\mathbf{s},\mathbf{y},\mathbf{t}}(\mathbf{x})$, we can write

$$\Delta\left(P_{\mathbf{s},\mathbf{y},\mathbf{t}}(\mathbf{x})\right) = \left(\Delta_{(\mathbf{y}\to\mathbf{t})}(\mathbf{x}),\Delta_{(\mathbf{s}\to\mathbf{y})}(\mathbf{x})\right)$$
$$= \frac{1}{\Delta_{(\mathbf{y}\to\mathbf{t})}(\mathbf{x})+\Delta_{(\mathbf{s}\to\mathbf{y})}(\mathbf{x})}(1-\tau,\tau).$$

Therefore $\Delta(\mathbf{z}) - \Delta(P_{\mathbf{s},\mathbf{y},\mathbf{t}}(\mathbf{x}))$ is orthogonal to (30).

### A.2.4 Close to an end-of-any-line

Close to the non-trivial end or start of any line, we don't have to be as careful with defining the displacement: any Lipschitz extension of the displacement we defined everywhere else would do, since here we are allowed (in fact, expect) to have fixed points.

For concreteness, let $(\mathbf{s} \to \mathbf{t})$ be the last Brouwer line segment in a path. In (26), we defined the displacement for points $\mathbf{x}$ such that $\sigma_{(\mathbf{s}\to\mathbf{t})}(\mathbf{x}) \leq 1 - \sqrt{h}$. For points such that $\sigma_{(\mathbf{s}\to\mathbf{t})}(\mathbf{x}) = 1$ (i.e. at the hyperplane through $\mathbf{t}$ and perpendicular to $(\mathbf{s}\to\mathbf{t})$), we simply set the default displacement $\hat{g}(\mathbf{x}) \triangleq \delta(\mathbf{0}_{3m},\mathbf{1}_m)$. For intermediate values of $\sigma_{(\mathbf{s}\to\mathbf{t})}(\mathbf{x}) \in [1-\sqrt{h},1]$, we simply interpolate according to $\sigma_{(\mathbf{s}\to\mathbf{t})}(\mathbf{x})$. Notice that this induces a fixed point for some intermediate point, since for $\mathbf{x}$ directly "above" the Brouwer line segment, $\delta\frac{\mathbf{z}-\mathbf{x}}{h}$ perfectly cancels $\delta(\mathbf{0}_{3m},\mathbf{1}_m)$. Define the displacement analogously at the (non-trivial) start of a path.

### A.2.5 Outside the picture

The displacement outside the picture is constructed by interpolating the displacement at $\frac{1}{m}\sum_{i=3m+1}^{4m} x_i = 1/2$, and the displacement at points in the "top" of the hypercube, where $x_i = 2$ for every $i$ in the last $m$ coordinates. The former displacement, where $\mathsf{E}_{i\in\{3m+1,\ldots,4m\}} x_i = 1/2$ is defined to match the displacement inside the picture. Namely, it is the default displacement everywhere except near the first Brouwer line segment which goes "down" from $\mathbf{s} = (\mathbf{0}_{3m}, 2\cdot\mathbf{1}_m)$ to $\mathbf{t} = (\mathbf{0}_{4m})$. Near this line, it is defined according to (26). (Notice that $\|\mathbf{t}-\mathbf{s}\|_2 = 1$.)

Formally, let $\mathbf{z}_{1/2} = \left(\mathbf{0}_{3m}, \frac{1}{2}\cdot\mathbf{1}_m\right)$; for $\mathbf{x}$ on the boundary of the picture, we have:

$$\hat{g}(\mathbf{x}) \triangleq \begin{cases} \delta(\mathbf{0}_{3m},-\mathbf{1}_m) & \|\mathbf{x}-\mathbf{z}_{1/2}\|_2 = 0 \\ \delta\frac{(\mathbf{z}_{1/2}-\mathbf{x})}{h} & \|\mathbf{x}-\mathbf{z}_{1/2}\|_2 = h \\ \delta(\mathbf{0}_{3m},\mathbf{1}_m) & \|\mathbf{x}-\mathbf{z}_{1/2}\|_2 \geq 2h \end{cases} \quad (31)$$

For points $\mathbf{x}$ such that $\sum_{i=3m+1}^{4m} x_i$ is very close to 2, the displacement $\delta(\mathbf{0}_{3m},\mathbf{1}_m)$ is not helpful because it points outside the hypercube, i.e. it would get completely erased by the truncation. Instead, we define the displacement as follows:

$$\hat{g}(\mathbf{x}) \triangleq \begin{cases} \delta(\mathbf{0}_{3m},-\mathbf{1}_m) & \|\mathbf{x}-\mathbf{z}_2\|_2 = 0 \\ \delta\frac{(\mathbf{z}_1-\mathbf{x})}{h} & \|\mathbf{x}-\mathbf{z}_2\|_2 \geq h, \end{cases} \quad (32)$$

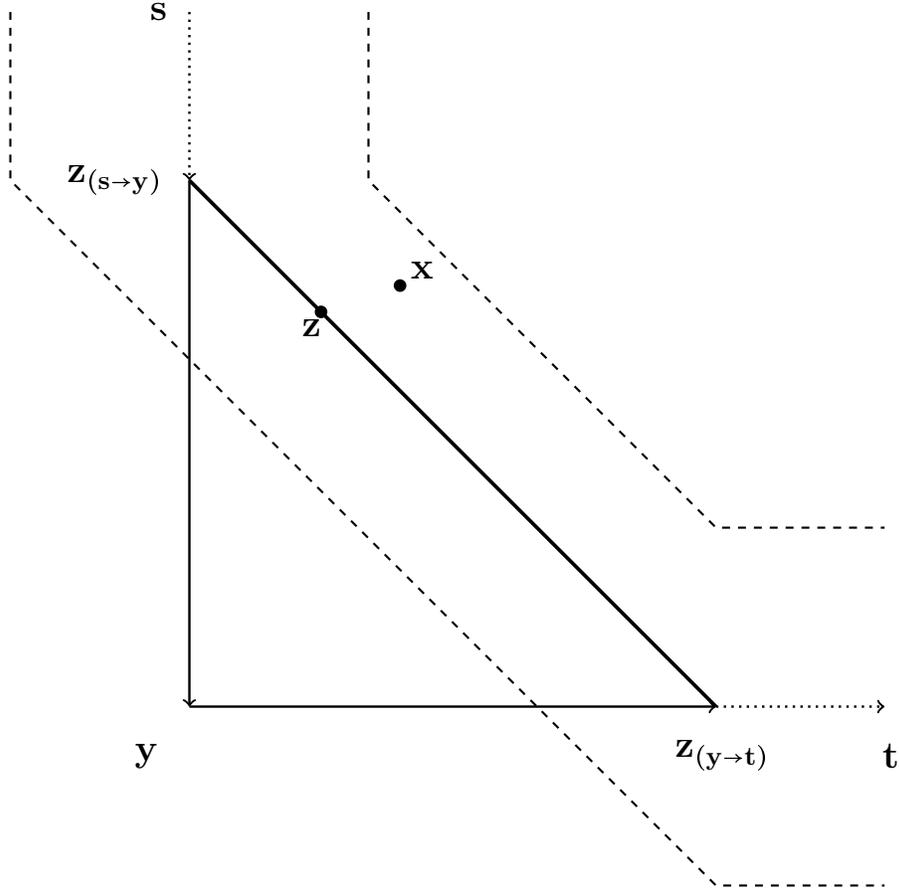

Figure 1: Geometry near a Brouwer vertex

The figure (not drawn to scale) shows some of the important points near a Brouwer vertex $\mathbf{y}$: There is an incoming Brouwer line segment from $\mathbf{s}$ through $\mathbf{z}_{(\mathbf{s}\to\mathbf{y})}$, and an outgoing Brouwer line segment to $\mathbf{t}$ through $\mathbf{z}_{(\mathbf{y}\to\mathbf{t})}$. For each point $\mathbf{x}$ between the dashed lines, we assign a point $\mathbf{z}$ on the line $L_{\mathbf{y}}$ as in (27), and define the displacement according to (28). Outside the dashed lines (including at $\mathbf{y}$ itself), we use the default displacement $\delta\left(\mathbf{0}_{3m}, \mathbf{1}_m\right)$.

where $\mathbf{z}_2 = (\mathbf{0}_{3m}, 2 \cdot \mathbf{1}_m)$. When $\theta \triangleq \sum_{i=3m+1}^{4m} x_i \in (1/2, 2)$, we interpolate between (31) and (32) according to $\frac{\theta - 1/2}{3/2}$.

## A.3 The class of local functions $\{f_{I_1, I_2}\}$

We start with defining two functions $f_{v', w'}$ and $f_{u', v', w'}$ which will be used in the definition of the class $\{f_{I_1, I_2}\}$.

Given $v', w'$ such that $(v', w')$ is an edge of the directed graph $G$, for $\mathbf{x}$ that is $2\sqrt{h}$-close to exactly one embedded edge $(v', w')$, we consider the definition of $f$ in the case where $(v', w')$ is an edge of the line, and we define $f_{v', w'}$ using the truncated displacement of Equations (26) and (28) depending on whether the point is close to one or two Brouwer line segments. For completeness of the definition (although it will not be used), we also define $f_{v', w'}$ at other points $\mathbf{x}$ inside the picture using the truncated default displacement. And for $\mathbf{x}$ outside the picture we define $f_{v', w'}$ in the same way we have defined $f$ (see Section A.2.5). We can think of $f_{v', w'}$ as "the function $f$ in case the line goes through $(v', w')$".

Given $u', v', w'$ such that $(u', v')$ and $(v', w')$ are edges of the directed graph $G$, for $\mathbf{x}$ that is $2\sqrt{h}$-close to the embedded vertex $\mathbf{x}^1(v', w')$, we consider the definition of $f$ in the case where $(u', v')$ and $(v', w')$ are edges of the line, and we define $f_{u', v', w'}$ using the truncated displacement of Equation (28). For completeness of the definition (although it will not be used), we also define $f_{u', v', w'}$ at other points $\mathbf{x}$ as we did for the function $f_{v', w'}$. We can think of $f_{u', v', w'}$ as "the function $f$ in case the line goes through $(u', v')$ and $(v', w')$".

Now we define the class $\{f_{I_1, I_2}\}$:

- For $\mathbf{x}$ outside the picture and every $I_1, I_2$, we define $f_{I_1, I_2}(\mathbf{x})$ as in Subsection A.2.5, which does not depend on the line.

- For $\mathbf{x}$ that is $2\sqrt{h}$-far from all Brouwer line segments, and every $(I_1, I_2)$, we define $f_{I_1, I_2}(\mathbf{x})$ according to the default displacement (which coincides with the definition of $f$).

- For $\mathbf{x}$ that is $2\sqrt{h}$-close to exactly one embedded edge from $v'$ to $w'$ the definition is as follows. For $i = 1, 2$, let $I_i = (T_i, S_i, P_i)$ be the local information of two vertices. In this region we define $f_{I_1, I_2}$ by cases.

    - If the first $m$-tuple of coordinates is $8\sqrt{h}$-close to $E(v')$ or $E(w')$ but $T_1 = 0$, we define $f_{I_1, I_2}(\mathbf{x})$ using the truncated default displacement.
    - If the first $m$-tuple of coordinates is $8\sqrt{h}$-close to $E(v')$ but the $S_1$-successor of $v'$ is not $w'$ (recall that each vertex has two possible successors the 0-successor and the 1-successor) we define $f_{I_1, I_2}(\mathbf{x})$ using the truncated default displacement.
    - If the first $m$-tuple of coordinates is $8\sqrt{h}$-close to $E(w')$ but the $P_1$-predecessor of $w'$ is not $v'$ we define $f_{I_1, I_2}(\mathbf{x})$ using the truncated default displacement.
    - For the second $m$-tuple of coordinates we define similarly but with respect to $I_2 = (T_2, S_2, P_2)$.
    - In all remaining cases we define $f_{I_1, I_2}(\mathbf{x}) = f_{v', w'}(\mathbf{x})$.

- In the remaining case, $\mathbf{x}$ is $2\sqrt{h}$-close to two embedded edges, therefore it has to be $2\sqrt{2}\sqrt{h} < 3\sqrt{h}$-close to an embedded vertex $v$ (i.e., $\mathbf{x}^1 = (E(v), E(v), \mathbf{0}_m, \mathbf{0}_m)$). This follows from the fact that any two consecutive Brouwer segments are orthogonal. Therefore, both the first and the second $m$-tuple of coordinates are $12\sqrt{h}$-close to encoding $E(v)$.

We denote by $u'$ the $P_1$-predecessor of $v$, and by $w'$ the $S_1$-successor of $v$. In this region we define $f_{I_1,I_2}$ as follows.

- If $T_1 = T_2 = 1$, $S_1 = S_2$ and $P_1 = P_2$, then we define $f_{I_1,I_2}(\mathbf{x}) = f_{u',v,w'}(\mathbf{x})$.
- Otherwise, we define $f_{I_1,I_2}(\mathbf{x})$ using the truncated default displacement.

It can be checked that indeed the properties (a)-(c) in the proposition are satisfied. Roughly speaking, this follows from the fact that we defined the function $f_{I_1,I_2}$ in a way that uses the information of $I_1$ in the case when the first $m$-tuple of coordinates in close to a valid vector, and it uses the information of $I_2$ in the case when the second $m$-tuple of coordinates in close to a valid vector. Regarding property (d), we first note that every point on any Brouwer line segment has either the first or the second $m$-tuple of coordinates a valid encoding of a vertex $E(v)$. This is because by the construction one of this $m$-tuples remains unchanged. Therefore, if $\mathbf{x}$ is $2\sqrt{h}$-close to such a line then either the first or the second $m$-tuple of coordinates is $8\sqrt{h}$-close to a valid encoding of a vertex. Hence, the last property simply follows from the fact that $f_{I_1,I_2}$ is defined using the default displacement in the case of a point that is far from all Brouwer line segments.